\documentclass[notoc]{JHEP} 
\usepackage{amsmath}
\usepackage{amssymb}
\usepackage{epsfig}
\usepackage{cite}
\psfull
\def\cA{{\cal{A}}}
\def\cN{{\cal{N}}}
\def\cB{{\cal{B}}}
\def\cC{{\cal{C}}}
\def\asb{{\bar \alpha}_{\mbox{\scriptsize s}}}
\def\as{\alpha_{\mbox{\scriptsize s}}}
\def\De{\Delta}
\def\ca{C_A}
\def\cf{C_F}
\def\nf{n_f}
\def\e{\mathrm{e}}
\def\SX{SMALLX}
\def\lepto{LEPTO}
\def\qth{{q_{t,h}}}

\def\etj{E_{t,jet}}
\def\cO#1{{\cal O}\left( {#1} \right)}
\def\vk{{\vec k}}
\def\vq{{\vec q}}
\def\vqt{{\vec q}_t}

\def\om{\omega}
\def\ebt{{\bar \eta}}
\def\qbt{{\bar q}_t}

\def\al{\alpha}
\def\mus{{\mu^{\ast}}}
\def\als{{\alpha^{\ast}}}

\def\kt{k_t}
\def\kq{(k\cdot q)}

\def\bfkl{{\scriptscriptstyle{BFKL}}}

\def\xf{x_f}
\def\xfb{\frac1z}
\def\xbf{z}

\def\Hbaro{\bar{H}^{(a)}}
\def\Hbard{\bar{H}^{(b)}}
\def\Hbarl{\bar{H}^{(c)}}

\def\ho{h^{(a)}}
\def\hd{h^{(b)}}
\def\hl{h^{(c)}}
\def\hm{h^{(d)}}

\newcommand\jpg[3]{{{\it J. Phys. }{\bf G #1} (#2) #3}}
\newcommand\cphys[3]{{{\it  Comp. Phys. }{\bf #1} (#2) #3}}
\newcommand\epjc[3]{{{\it  Eur. Phys. J. }{\bf C#1} (#2) #3}}

\title{Small-$\boldsymbol{x}$ one-particle-inclusive quantities in the
  CCFM approach\thanks{This work was supported in part by the EU
    Fourth Framework Programme `Training and Mobility of Researchers',
    Network `Quantum Chromodynamics and the Deep Structure of
    Elementary Particles', contract FMRX-CT98-0194 (DG 12-MIHT).}}
  \author{G. Bottazzi, G. Marchesini, G.P. Salam and M. Scorletti \\
    Dipartimento di Fisica, Universit\`a di Milano, and INFN, Sezione
    di
    Milano, Italy\\
    E-mail: \email{Giulio.Bottazzi@mi.infn.it},
    \email{Giuseppe.Marchesini@mi.infn.it},
    \email{Gavin.Salam@mi.infn.it},
    \email{Massimo.Scorletti@mi.infn.it}}
  
  \abstract{This article presents the results of a quantitative study
    of the small-$x$ data at HERA, using the CCFM equation.  The first
    step consists of choosing the version of the CCFM equation to be
    used, corresponding to selecting a particular subset of
    next-to-leading-logarithmic corrections --- the choice is
    constrained by requiring a phenomenologically reasonable small-$x$
    growth.  For the time being, the parts of the splitting functions
    that are finite at $z=0$ have been left out.  We then examine
    results for $F_2^c$, $R$, the transverse energy flow, the
    charged-particle transverse-momentum spectrum and the forward-jet
    cross section and compare to data.  While some of the data is
    reproduced better than with DGLAP-based calculations, the
    agreement is not entirely satisfactory, suggesting that the
    approach developed here is not yet suitable for detailed
    phenomenology. We discuss why, and suggest directions for future
    work.}

\preprint{hep-ph/9810546\\
          IFUM-634-FT\\
          October 1998}

\keywords{QCD, Deep Inelastic Scattering, Jets, Phenomenological Models}

\begin{document}

\section{Introduction}
\label{sec:intro}

The HERA accelerator has opened up a radically new region of phase
space for Deep Inelastic Scattering (DIS). This is the region where
Bjorken-$x$ is very small, while the photon virtuality $Q^2$ is still
hard but moderate.  In this kinematical domain it is no longer clear
that DGLAP evolution \cite{DGLAP}, which resums logarithms of $Q^2$,
should lead to a good description of the physics. Indeed one expects
that the resummation of logarithms of $1/x$ should become equally if
not more important. One of the aims of the HERA experiments is to
study observables which have different characteristics according to
the kind of resummation performed, in order to clarify one's
understanding of the kinematical domain and of the details of the
resummation of logarithms of $x$.

For total cross sections (i.e.\ DIS structure functions), leading
logarithms of $1/x$ ($\as^n\ln^n 1/x$) can be resummed using the BFKL
equation \cite{BFKL}.  This predicts that the total cross section
should grow as a power of $1/x$. The rapid rise of the $F_2$ structure
function at small $x$ \cite{F2H1,F2ZS}, when first discovered, was
optimistically hailed as being evidence for this. There are however
difficulties with such a claim at a quantitative level:
\begin{enumerate}
\item With DGLAP-based approaches the data can be quite
  straightforwardly fitted, using either an input parton distribution
  which rises at small-$x$, or by starting the evolution from a low
  scale \cite{MRST,GRV}.
\item Leading logarithmic (LL) BFKL evolution leads to much too steep
  a rise of $F_2$ to allow a fit to the data (see e.g.\ \cite{BF}).
\item The recently calculated next-to-leading (NLL) corrections to the
  BFKL kernel \cite{NLLBFKL,CCFL} turn out to dominate over the
  leading contribution for any realistic value of the strong coupling
  (so perhaps it should come as no surprise that leading-order BFKL
  cannot fit $F_2$), and if taken literally, lead to nonsensical
  results \cite{Ross}.
\item BFKL evolution is not ordered in transverse momentum, as a
  result of which diffusion in transverse momentum \cite{Bart,Muel,CC1}
  leads to significant dependence on the infra-red region, where the
  perturbative series is irretrievably out of control.
\end{enumerate}
Points two and three are closely connected, and in due time, perhaps
following an approach similar to that of \cite{salam}, we will
probably learn how to resum large parts of the NLL corrections. One
can to some extent avoid the problems of the infrared region by
including small-$x$ resummations in the $Q^2$-evolution of the $F_2$
structure function \cite{BFKLQ2}; the tradeoff is that as in DGLAP,
one needs an input parton distribution over the whole range of $x$,
possibly leaving a little too much freedom in fitting the data.

A way of observing BFKL phenomena while avoiding sensitivity to poorly
known input parton distributions or to the infrared region was
suggested some time ago by Mueller and Navelet \cite{MuNa} in their
proposal to study more exclusive quantities, in particular the cross
section for the production, in hadron-hadron collisions, of two
similarly hard jets separated by a large rapidity interval, which
should grow as an exponential of that rapidity interval. The advantage
of this observable (and of many other similarly exclusive measurements
which have since then been proposed), is that by triggering on two
large transverse momenta at the extremes of the evolution, one can
suppress the probability of diffusion into the infra-red region,
leaving a quantity which can, at least in principle, be calculated
perturbatively (of course if the rapidity interval between the two
transverse momenta becomes too large, then diffusion once again
becomes important). A second advantage is that such observables can
discriminate between BFKL and DGLAP evolution more effectively than
the total cross section at small $x$: DGLAP evolution cannot possibly
mimic the expected BFKL behaviour, because generally the two
transverse scales are of the same order, and DGLAP only enhances cross
sections for interaction between objects with very different
transverse scales.

So an efficient way to identify the most important characteristics
features of $\ln x$ resummation is to study the final state. It turns
out that the multi-parton distribution of an event cannot be reliably
obtained from the BFKL equation. The fundamental additional element
that must be taken into account (as in almost any calculation of
final-state properties \cite{IR}) is that of the coherence of QCD
radiation \cite{DKT}. For the particular case of an evolution in which
the transverse scale increases along the direction of evolution, the
correct formulation was derived in \cite{CCFM} and is known as the
CCFM equation --- it correctly resums the leading $\ln 1/x$ and $\ln
Q$ terms both for inclusive and non-inclusive quantities \cite{GM}.
Angular ordering should also reduce the dependence on the infra-red
region, since large drops in the transverse scale are disfavoured,
thus limiting diffusion \cite{BMSS}.

In contrast to the BFKL and DGLAP equations, whose parton densities
depend on two variables (energy fraction and transverse momentum), the
CCFM gluon density depends additionally on a limiting angle. This
increases very considerably the mathematical complexity of the
equation, and it can only be solved numerically (and even that is
highly non-trivial).

In view of this difficulty one is led to ask whether it is really
necessary to use the CCFM equation in order to get a valid description
of the final state. It has been pointed out by Ryskin \cite{RyskPriv}
that if one formulates the BFKL equation as an evolution in rapidity
rather than in $x$, then one automatically has angular ordering. This
for example is the approach that has been taken by Orr and Stirling
\cite{OrrStir} in calculating Mueller-Navelet-like two-jet observables
at the Tevatron, and in such a context it may be valid.

However the point of the CCFM equation is not simply that of angular
ordering --- rather it consists in the \emph{combination} of the
ordering of angles and the ordering of the $x$ variables, in a
situation in which transverse momenta increase along the chain, i.e.\ 
a DIS-type situation. If one applies the BFKL equation with only
rapidity ordering to a DIS situation then one will obtain an answer
for the cross section with spurious double transverse logarithms
($\as^2\ln1/x \ln^2 Q$) at NLL order. This problem is closely
connected with that of the choice of scale in the kernel
\cite{CCFL,salam} --- indeed in the language of \cite{CCFL}, the
BFKL equation with rapidity ordering corresponds to choosing a
symmetric scale $k_1 k_2$ in the LL kernel. What this means in
practice is that the cross section at small $x$ and large $Q^2$,
instead of behaving, as it should do, as
\begin{equation*}
  \exp\left( 2\sqrt{\frac{\asb \ca}{\pi} \ln \frac1x \,\ln Q^2}
  \right)\,, 
\end{equation*}
goes as 
\begin{equation*}
  \exp\left( 2\sqrt{\frac{\asb \ca}{\pi} \ln \frac{Q}{x} \,\ln Q^2}
  \right)\,. 
\end{equation*}
It turns out though, that even the BFKL equation with ordering in $x$
(which does not suffer from the above problem) manages to reproduce
correctly certain properties of the CCFM final state: in a fixed-order
calculation by Forshaw and Sabio Vera \cite{FSv}, subsequently
extended to all orders by Webber \cite{Webb} it was demonstrated that
in the double-logarithmic approximation all one-particle inclusive
quantities have the same leading-logarithmic terms in both approaches.
Nevertheless one should be wary of such a result, since there is the
distinct possibility of one-particle-inclusive quantities having a
collinear divergence at subleading order in BFKL \cite{GPSfuture}, and
secondly, BFKL may not correctly reproduce correlations between
emissions \cite{CMWfuture,GPSfuture}.

So it is likely that the CCFM equation is the best approach currently
available for studying the final state. At the same time it should be
noted that the CCFM equation is not entirely free of problems. Its
approximation of ordering simultaneously in angles and $x$ is valid
when the transverse scales increase along the direction of evolution.
When the transverse scales decrease, the relevant approximation is that
of ordering simultaneously in angles and $\bar x$.  One of the
characteristics of small-$x$ evolution is that even if, predominantly,
the transverse scale increases (as in DIS), there are evolution steps
in which it drops (diffusion) --- and the CCFM equation is subject to
large subleading corrections in those steps (again this is related to
the issue of the choice of scales).  One might be tempted to argue
that in DIS the problem arises only occasionally and so will not
matter excessively --- but several final-state ``BFKL-signals'' at
HERA tend to favour evolution paths in which the evolution has no
dominant direction in transverse momentum.

Other problems that arise are connected with two further subleading
elements of the CCFM equation: the treatment of factors of $(1-z)$ in
the kinematical variables of the virtual corrections is not defined at
LL order, and as we will see, a modification of their treatment can
lead to very large subleading corrections. Secondly the part of the
splitting function which is finite as $z\to0$,
\begin{equation*}
-2+ z(1-z) + \frac{1}{1-z}  
\end{equation*}
can also give important subleading corrections.\footnote{Here the
  $z\to1$ singularity is regulated by a corresponding Sudakov form
  factor.} There are several possible ways of including it in the
evolution, and they may well give significantly different subleading
corrections. Because of the enhancement at $z=1$, these contributions
are quite difficult to study numerically. It should be noted that they
(in particular the $1/(1-z)$ part) are vital for a correct description
of infra-red sensitive final-state properties.

In this uncertain theoretical context, one thing is clear however:
experimentally, DGLAP-based approaches (whether Monte-Carlo event
generators or NLO calculations) are incapable of describing many of
the small-$x$ final-state observables that are supposed to be
sensitive to BFKL phenomena \cite{H1Et,H1kts,ZEUSfj,H1fj96}.  So there
is an urgent need for some BFKL-based phenomenology, and in particular
a Monte Carlo event generator based on a leading $\ln x$ resummation.
This article is in some sense intended as a study of the extent to
which one can carry out quantitative phenomenology using the CCFM
equation as it stands currently.

At this point, it is worthwhile specifying some realistic aims. Quite
a significant issue is whether one should worry about matching the NLL
corrections calculated for the BFKL equation \cite{NLLBFKL,CCFL}. If
they were small, the answer would obviously be ``yes.'' But since they
are so large, all that should matter in practice is the resummed
corrections. Some idea of what a (partially) resummed result might
look like for the BFKL equation is given in \cite{salam}. Beyond this
we are limited to being guided by the experimental data. In
particular, we aim to reproduce the total cross section as a function
of $x$ and $Q^2$. This will force us to deal with problems associated
with the treatment of the infrared region.

Bearing in mind these requirements, our approach is as follows: we
choose a CCFM evolution equation which phenomenologically seems to
have a reasonable exponent (i.e.\ growth at small $x$). By
``phenomenologically reasonable,'' we mean that for the relevant
values of $\as$ it should be roughly consistent with the growth of
$F_2$, and with expectations from a resummation of NLL terms of the
BFKL kernel (as for example, at an approximate level, in
\cite{salam}).  The methods used to determine the exponent follow on
from those of \cite{BMSS}. Because of the lack of a well-defined way
of including the $z\to0$ finite part of the splitting function, and
because of the considerable difficulty in implementing each new way
that one might invent, we simply leave it aside, for possible
inclusion at a later stage.

We then develop a suite of programs allowing us to study structure
functions and one-particle inclusive quantities, and constrain the
treatment of the infra-red region by fitting to the $F_2$ structure
function data.  At this point we should be in a position to give
relatively parameter-free predictions for other structure functions
($F_2^c$, $F_L$) and various (almost) one-particle-inclusive
final-state quantities: the transverse energy flow, charged particle
spectra and the forward-jet cross section.

Before entering into the details it should be noted that there are
already in existence a number of programs based on the CCFM equation,
for studying small-$x$ DIS scattering. 

In \cite{KMSccfm} there have been studies of the $F_2$, $F_2^c$ and
$F_L$ structure functions using the form of the CCFM equation without
the $z\to0$ finite part of the splitting function, and with all other
$1-z$ factors replaced by $1$ (both in the real and virtual
corrections).

The program \SX\ \cite{SMALLX} is a direct Monte Carlo implementation
of the CCFM equation. It includes the $z\to0$ finite part of the
splitting function (unlike the results presented here), and
additionally, the exact kinematic constraints.  The main problem with
this program is that it is a forward-evolution event generator, with
weights that have extremely large fluctuations.  This renders it
somewhat unwieldy to use. Additionally, though it contains all the
physics, there is no way of modifying the exact implementation of the
subleading corrections --- such modifications turn out to have large
effects. The \SX\ program has recently been used and studied in
\cite{Biernat}, for the case of the production of charm quarks ---
unfortunately the precision of the data for charm quark production is
rather poor (the tagging of open charm is not a simple task), and
there is little final-state data available.

Another major numerical study is based on the Linked Dipole Chain
(LDC) model \cite{LDCorig}, which has been implemented as a Monte
Carlo event generator \cite{LDCmc}.  Formally it is based on the CCFM
equation with some simplifications and modifications. Perhaps the most
important modification is that the evolution has been made symmetric
(i.e. it is equally valid whether the transverse scale increases or
decreases along the direction of evolution). A lack of symmetry is one
of the main defects of the CCFM equation. On the other hand, the LDC
model has the problem that it rearranges the real and virtual terms of
the evolution in a manner which alters the LL small-$x$ evolution.
Whether this is a serious problem is to some extent a question of
opinion --- after all, given that the known NLL corrections are so
large, what really matters phenomenologically is the resummed
evolution, not its perturbative expansion. The LDC approach gives
satisfactory results for the structure function data and some final
state properties such as the transverse energy flow. However it has
problems with the charged-particle transverse momentum spectrum and
with the forward-jet cross section. As will be seen, our approach has
rather similar problems --- a discussion of this will be given in the
conclusions.

The structure of this paper is as follows: in section~\ref{sec:ao} we
review the structure of the CCFM equation, discuss approximations that
are commonly made when solving it, together with the treatment of
subleading corrections in the non-Sudakov form factor. In
section~\ref{sec:expn} we examine the asymptotic exponents of various
versions of the CCFM equation. In section~\ref{sec:sf} we consider the
evolution parameters and corresponding fits to the $F_2$ structure
function, as well as the consequences of these different fits for
$F_2^c$ and $R$. In section~\ref{sec:1pi} we then discuss the
determination of one-particle-inclusive final-state quantities, and
present results. The whole approach, and directions for future work
are finally examined in the light of these results in
section~\ref{sec:conc}.

In an appendix we give details of a novel analytic approach for
implementing the off-shell boson-gluon fusion
\cite{CCH,CollinsEllis,LevinEtAl}, which is convolved with the
unintegrated gluon structure function to give physical cross sections.
Traditional approaches generally require a two-dimensional numerical
integration over the quark-antiquark phase-space (or alternatively a
treatment in terms of the Mellin transform) --- the solution presented
here carries out these integrations analytically, even in the presence
of massive quarks, greatly simplifying the determination of physical
cross sections from the unintegrated gluon structure function.

\section{Angular ordering}
\label{sec:ao}

\subsection{Kinematics}
\label{sec:kine}
In what follows, we discuss the unintegrated gluon density of a proton
(taken to be massless) with energy $E_p$. This gluon density,
$\cA(x,k,p)$, depends on: $x$, the fraction of the proton's
plus-momentum component that is carried by the gluon; $k$ the gluon's
transverse momentum; and $p$, which defines the maximum-allowed angle
of all emissions prior to this gluon, through the relation $p = 2xE_p
\tan \theta_{\max}/2$, where $E_p$ is the proton energy. The recipe
for choosing $p$ in DIS will be discussed later.

The angular ordering condition on the last emitted gluon (with
transverse momentum $q_t$ and plus-momentum fraction $x(1/z - 1)$) can
be written as
$$
p > \frac{zq_t}{1-z} \,=\, z q\,,
$$
where for convenience we have introduced the variable $q =
q_t/(1-z)$. We shall also use the rapidity $\eta$ of an emitted
particle:
\begin{equation}
  \label{eq:rap}
  \eta = \frac12 \ln \frac{q_-}{q_+} = \ln \frac{zq}{2xE_p}\,,
\end{equation}
where, $q_\pm = q_0 \pm q_3$ and the proton has a positive
3-component. For the purpose of the kinematics, emitted gluons will
always be taken to be massless.

\begin{figure}
\begin{center}
\epsfig{file=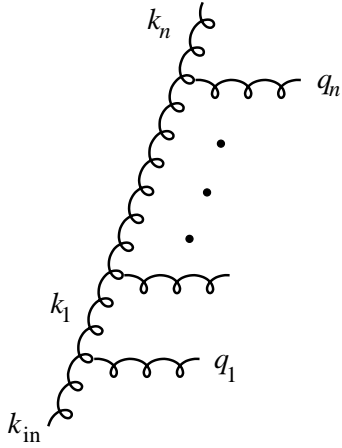}
\end{center}
\caption[]{Kinematics of the branching process: $k_{in}$ is momentum
coming from proton; $k_n$ is momentum entering into the photon-gluon
fusion quark-box.}
\label{fig:kin}
\end{figure}

Considering a general branching (see figure \ref{fig:kin}), in which
a space-like gluon $k_n$ has transverse momentum $\vk_n$ and momentum
fraction $x_n$, and an emitted gluon $q_n$ has transverse momentum
$\vq_{t,n}$, then the general ordering is $q_{n+1} > z_n q_n$ (with
$z_n=x_n/x_{n-1}$ and $\vq_n = \vq_{t,n}/(1-z_n)$), with $\vk_n =
\vk_{n-1} - (1-z_n)\vq_n$. Additionally all the $q_{t,n}$ satisfy
$q_{t,n} > \mu$ where $\mu$ is our collinear cutoff.

The $1-z_n$ factor is necessary for the exact reconstruction of the
transverse momenta (alternatively, one can interpret it as being
necessary for exact angular ordering). It gives a next-to-leading
contribution to the structure function evolution (as does angular
ordering itself).

\subsection{The CCFM equation}
\label{sec:ccfm}

The equation satisfied by the unintegrated gluon structure function,
$\cA(x,k,p)$ is 
\begin{multline}
  \label{eq:ccfm}
  \cA(x,k,p) = \cA^{(0)}(x,k,p) + \\
  \int_x^1 \frac{dz}{z} \int \frac{d^2q}{\pi q^2} \,\asb\left(q_t\right)
  \;\cA(x/z,|\vk+\vq_t\,|,q)\;
  \De(z,k,q) \Theta(p-zq)\,,
\end{multline}
where $\asb=\as\ca/\pi$ and $\vq_t = (1-z)\vq$.  The non-Sudakov form
factor (virtual correction) is
\begin{equation}
  \label{eq:nsff}
  \ln \De(x,k,q) = -\int_x^1  \frac{dz}{z} \int \frac{d^2q'}{\pi
    {q'}^2}\, 
  \asb({q_t}') \, \Theta(k - T q')\, \Theta(q'-zq)\,.
\end{equation}
It is usual to set $T=1$, however at leading-logarithmic order it is
equally valid to choose $T=1-z$. A motivation for choosing the latter
form is that in the real emission term, one has a dependence on
$\vk+(1-z)\vq = \vk+\vq_t$; one can argue that in analogy, since $q'$
is a rescaled transverse momentum, $q' = q'_t/(1-z)$, it is
$\Theta(k-q_t')$ and not the more usual $\Theta(k-q')$ which should
appear in the form factor. More formally the choice $T=1-z$, at first
sight, looks like it ought to cancel the contribution from all
emissions with $q_t < k$.

The two choices will be discussed in more detail in
section~\ref{sec:expn}.

\subsection{The initial condition}
\label{sec:initcond}

To ensure a sensible initial condition (one which leads to a collinear
safe result after evolution), $\cA^{(0)}$ should correspond to the
distribution for a Reggeised gluon. Relating to a normal,
``non-Reggeised,'' initial condition, $\cA ^{(\mathrm{init})}$, we
have
\begin{equation}
  \cA^{(0)}(x,k,p) = \int_z^1 \frac{dz}{z}\,
  \frac{d\cA^{(\mathrm{init})}(z,k)}{d\ln 1/z} \,\De(x/z,k,\mu) \,.
\end{equation}
The form that we take for $\cA^{(\mathrm{init})}(x,k)$ is
\begin{equation}
  \label{eq:Ainit}
  \cA^{(\mathrm{init})}(x,k) = 
  \cN \frac{n+1}{\pi k_0^2}\, (1-x)^{n}\, \e^{-k^2/k_0^2}\,,
\end{equation}
with $\cN$, $n$ and $k_0$ parameters to be discussed below. Note that
$\cA^{(0)}(x,k,p)$ does not depend on $p$.  The
normalisation is such that
$$
xg(x,Q^2) = \int d^2k \,\cA(x,k,Q)\,,
$$
where $g(x,Q^2)$ is the gluon density in the proton; $\cN$ is the
result of the energy sum rule for the initial condition:
\begin{equation}
\cN = \int d^2k dx \cA^{(\mathrm{init})}(x,k)\,.
\end{equation}

\subsection{The running coupling}
We use the one-loop running coupling, taking $q_t$ as its scale (in
accord with the results obtained from the NLL calculation
\cite{CCFL}). Since $q_t$ kinematically can go down to small values,
it is necessary to introduce a prescription for the running of $\as$
at low scales: we modify $\as$ so that it reaches a plateau in the
infra-red region:
\begin{equation}
  \label{eq:alpha}
  \as(q_t) = \frac{1}{b_0 \ln (q_t^2 + \Lambda_0^2)/\Lambda^2}\,,
\end{equation}
with 
$$
b_0 = \frac{11\ca - 2\nf}{12\pi}\,,
$$
and $\Lambda_0$ chosen so as to ensure that $\as(0)=\as^{(0)}$,
with $\as^{(0)}$ a parameter; finally, we take $\nf=4$ and $\Lambda =
0.135$~GeV.

\subsection{Parameters}
To summarise, the parameters that must be chosen are the following:
$\as^{(0)}$, the value of $\as$ at which it freezes, a cutoff on
emitted transverse momenta $q_t > \mu$ (and $q_t'>\mu$ in the virtual
corrections), and the parameters which affect the properties of the
initial condition, $\cN$, $n$ and $k_0$. They will be constrained by
requiring a good fit to the $F_2$ structure function, as discussed in
section~\ref{sec:sf}. One consistency check of our approach will be
that it is possible to find a set of fit parameters that is in accord
with one's physical expectations. For example the normalisation should
be such that the energy sum rule should integrate to $\cN\simeq0.5$;
$k_0$ is expected to be of the order of $1$~GeV; $\as^{(0)}$ should be
of the order of $0.5$; and $n\simeq 4$.

Parameters that are only weakly constrained by $F_2$ should then be
varied to probe their effect on final state quantities.

\section{Asymptotic exponents}
\label{sec:expn}
\begin{figure}
  \begin{center}
    \epsfig{file=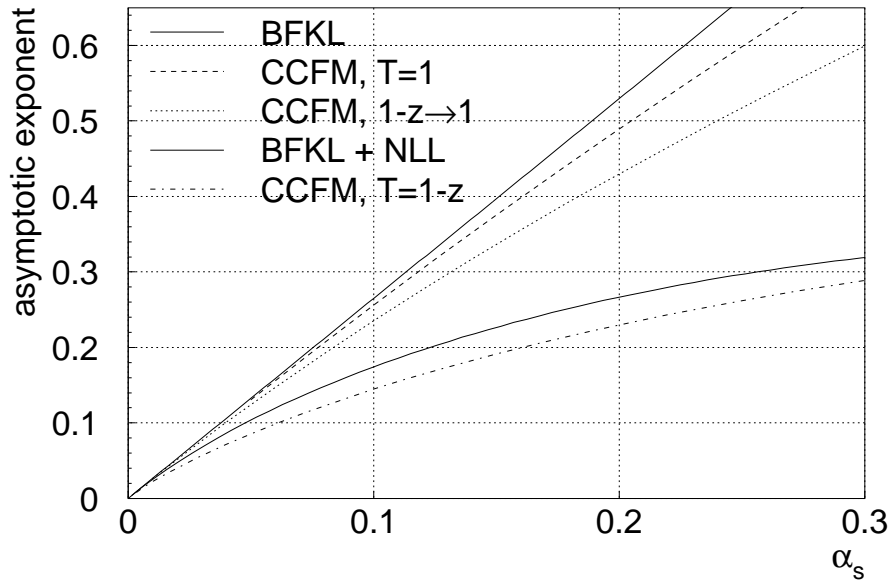,width=0.75\textwidth}
    \caption[]{The asymptotic exponent of the BFKL equation and
      various forms of the CCFM equation, as a function of $\as$. The
      ``BFKL $+$ NLL'' curve is the resummed result (scheme 3) of
      \cite{salam}. }
    \label{fig:expn}
  \end{center}
\end{figure}
Before engaging in a fit of the $F_2$ structure function, it is
helpful to study, as a function of fixed $\as$, the ``asymptotic
exponents,'' $\omega$, of the various forms of the CCFM equation,
since this will indicate which one is most likely to be in accord with
the phenomenology (though it should be borne in mind that $\om(\as)$
is not physically observable).  The technique adopted is based on that
developed in \cite{BMSS}:
\begin{equation}
\om = \frac{d}{d \ln x} \ln A(x,k,p)\,,
\end{equation}
with $x$ sufficiently small that the result is independent of $x$,
$k$, and $p$. In the BFKL case, $\om=\asb 4\ln2$.

Figure~\ref{fig:expn} shows the asymptotic exponents of several
versions of the CCFM equation, compared to the BFKL equation, as a
function of $\as$. The three versions of the CCFM equation are as
follows: the case ``$1-z\to1$'' corresponds to (\ref{eq:ccfm}) and
(\ref{eq:nsff}) with all $1-z$ factors replaced by $1$ (also in the
relation between $q$ and $q_t$).  This is an approximation quite
commonly used (studied for example in some detail in \cite{BMSS})
because it greatly simplifies the numerical solution.  However it
leads to a slight violation of the angular ordering (or to an
incorrect reconstruction of the transverse momenta, depending on
whether one interprets $q$ as a real, or a rescaled transverse
momentum). The other two versions of the CCFM equation implement the
correct treatment of angular-ordering (or transverse-momentum
reconstruction), but differ in their treatments of the form factor.
The $T=1$ case corresponds to the version of the CCFM equation more
commonly quoted.

One sees immediately that the $1-z\to1$ and $T=1$ cases have exponents
which are rather similar to that of the leading-logarithmic BFKL
equation. In particular one sees that for the values of $\as$ which
are typical in HERA physics, namely $0.2$--$0.3$, the exponents are
far above anything observed experimentally. In a non-asymptotic
situation (i.e.\ $x$ not infinitely small) these exponents can be
reduced by the introduction of a large collinear cutoff, but, it turns
out, not sufficiently to allow a fit to the data.

So the version of the CCFM equation that we will concentrate on in the
remainder of this article will be that with $T=1-z$. One can calculate
numerically its next-to-leading correction to the exponent, which
turns out to be extremely large: $\omega = 4\ln2 \asb - (75\pm4)
\asb^2 + \cO{\as^3}$.  This is to be compared with $\omega = 4 \ln 2
\asb - 17.91 \asb^2$ in the case of the full NLL BFKL corrections
($\nf=0$) \cite{NLLBFKL,CCFL}. One is immediately induced to ask why
setting $T=1-z$ produes such a large correction.  Qualitatively, it
leads to large negative virtual corrections from $q_t' \lesssim k$ for
$z$ close to $1$. However the corresponding real-emission
contribution, $q_t \lesssim k$ for $z$ close to $1$, turns out to be
dynamically suppressed because it corresponds to a large emission
angle, which reduces the phase space available for subsequent
emissions. Consequently the asymptotic exponent receives a large
negative correction.

Another question is whether it matters that the NLL correction is so
much larger than the true one. One can argue not, since the pure NLL
correction turns out to be visible only for extremely small values of
$\as$, while phenomenologically we are interested in the region
$\as\simeq0.2$, where formally more subleading corrections (e.g.\ NNLL
etc.) are just as important. So what matters should be the resummed
asymptotic exponent, and from figure~\ref{fig:expn} it seems that this
quantity is roughly in line with phenomenological expectations. In
fact if one carries out a partial resummation of subleading
corrections to the exact NLL BFKL exponent (the curve shown in
figure~\ref{fig:expn} corresponds to scheme 3 of\cite{salam}), one
obtains a result which is not so different from that with the equation
being used here.

Having completed the discussion of the asymptotic exponent, it is
however vital to remember that much of the HERA data is to be found in
regions of $x$ (or energy) which may be quite far from asymptotic.

\section{Structure functions}
\label{sec:sf}

\subsection{Convolution with the quark box}
\label{sec:hme}

To obtain a result for the $F_2$ structure function (or any other
cross section) it is necessary to perform the convolution of the
unintegrated gluon density $\cA(x,k,p)$ with the boson-gluon fusion
matrix element (which is discussed in the appendix):
\begin{equation}
  \label{eq:bgfconv}
  F_2(x,Q^2) = \sum_q e^2_q \int_x^1 \frac{dz}{z} \int \frac{dk}{k}
  {\hat\sigma}_{F_2} (z,k^2,Q^2,m^2_q) \cA(x/z,k,p)\,.
\end{equation}
In a similar way, one can determine $F_L$ and $F_2^c$. The maximum
angle is taken, as in the program \SX\ \cite{SMALLX}, to be defined by
the rapidity of the combined $q\bar{q}$ pair
$$
p^2 = \frac{Q^2}{z(1-z)}\,.
$$
Modifying this limit would alter our results at next-to-leading
order. The light quarks are taken to be massless, while for the heavy
quarks we use $m_c = 1.5$~GeV and $m_b = 5.0$~GeV. The convolution is
performed in a frame in which the proton and photon are collinear and
along the $z$-axis (this differs from the laboratory frame by a
rotation and a boost). Correspondingly, transverse momenta for
final-state quantities are also determined in such a frame.

As discussed in the appendix, $\hat \sigma$ is determined entirely
analytically. This is in contrast to other approaches, which generally
leave $\hat \sigma$ in the form of a double integral which must be
evaluated numerically \cite{CCH,CollinsEllis,LevinEtAl,BW,KLM}. The
reduction in the number of numerical integrations simplifies
considerably the implementation of the convolution. The only
disadvantage is that the analytic integration must be performed with
fixed $\as$ --- so the value that we use is given by the logarithmic
mean of $\as$ between scales $\sqrt{k^2+m^2_q}$ and
$\sqrt{Q^2+m^2_q}$:
\begin{equation}
  \langle \as \rangle = 2 
  \left(\ln \frac{Q^2+m^2_q}{k^2+m^2_q}\right)^{-1}
  \int_{\sqrt{k^2+m^2_q}}^{\sqrt{Q^2+m^2_q}} \frac{dq}{q} \,\as(q)\,.
\end{equation}
In the limits $k\ll Q$ and $k\gg Q$ this gives the correct overall
dependence on the running coupling (possibly incorrect with respect to
details of the treatment of the quark mass).

\subsection{Structure function results}
\label{sec:sfres}

\begin{figure}
  \begin{center}
    \epsfig{file=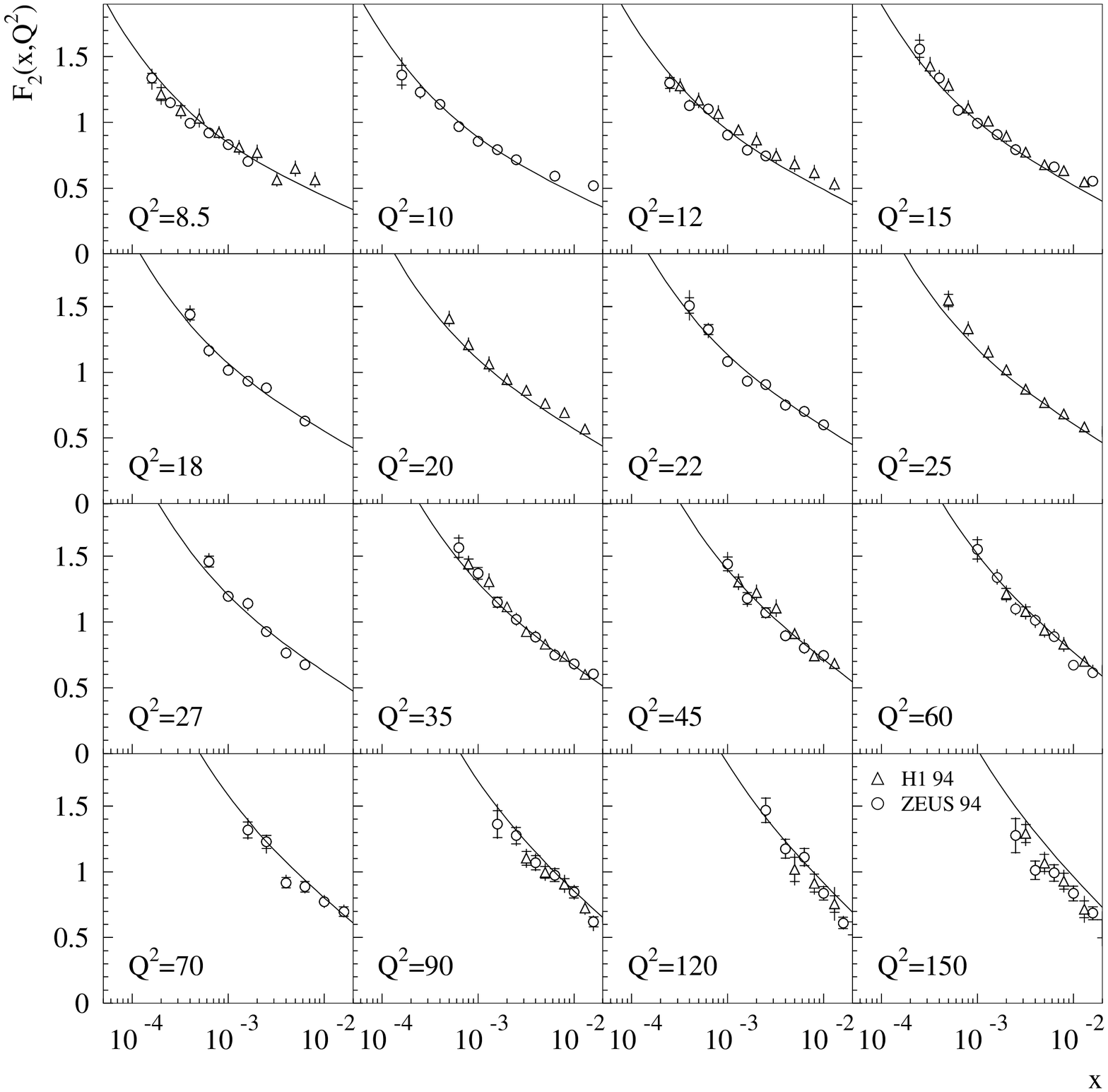,width=0.8\textwidth}
  \end{center}
  \caption{Fits to the $F_2$ structure function, compared with H1
    \cite{F2H1} and ZEUS \cite{F2ZS} 1994 data points.}
  \label{fig:f2}
\end{figure}
The fit for the structure function is performed in the kinematic range
$x<10^{-2}$ and $8 < Q^2 < 140$~GeV$^2$.  The parameters to which the
fit is most sensitive are $\mu$, the collinear cutoff, and $\alpha_0$,
the value at which $\as$ freezes.  We find that $0.01\lesssim
\mu\lesssim0.1$~GeV, and $\alpha_0\sim0.5$ lead to reasonable fits of
the H1 and ZEUS 1994 structure function data \cite{F2H1,F2ZS}; as one
would expect, there is relatively little dependence on the value of
$n$ (the power of $(1-x)$ in the initial condition, see
\eqref{eq:Ainit}), though it does affect the fitted value of $\cN$.
It turns out that for the $k$-dependence of the initial condition,
$\propto e^{-k^2/k_0^2}$, there is a correlation between the value of
$k_0$ and the optimal value of $\mu$. Good fits are obtained with a
range of $k_0$: we will use $k_0 = 1$~GeV, and then examine the effect
on other quantities of varying it.  Modifying $k_0$ by a factor of $2$
(and changing $\mu$ and $\alpha_0$ accordingly) affects the
normalisation by about $50\%$.

The set of parameters that will be used is as follows:
\begin{center}
  \begin{tabular}{|c|c|}
    \hline
    $\mu$ & $0.075$~GeV \\ \hline
    $\as^{(0)}$ & $0.60$ \\ \hline
    $k_0$     & $1.0$~GeV \\ \hline
    $n$  & $4$ \\ \hline
    $\cN$  & $0.72$ \\ \hline
  \end{tabular}
\end{center}
A comparison with the data is shown in figure~\ref{fig:f2}, and is
seen to be in good agreement. The $\chi^2$ are $64.4$ for $59$ H1
points \cite{F2H1} and $84.7$ for $80$ ZEUS points \cite{F2ZS}.

\begin{figure}
  \begin{center}
    \epsfig{file=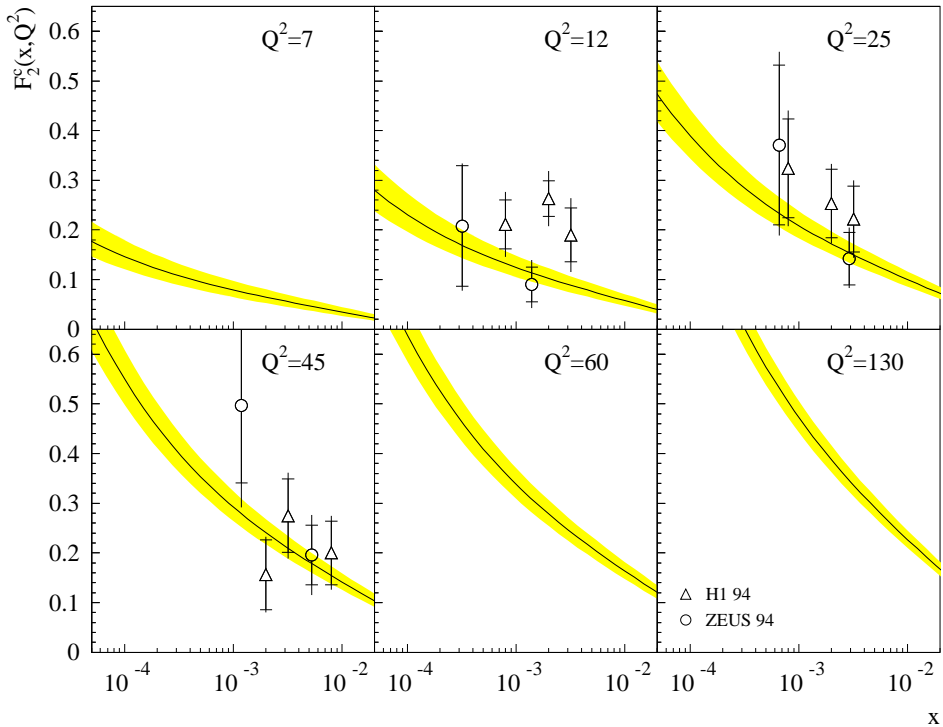,width=0.8\textwidth}\\
  \end{center}
  \caption[]{Results for $F^{c}_2$ compared with H1 1994 data
    \cite{F2cH194}, and ZEUS 1994 data \cite{F2cZS94}.}
  \label{fig:f2c}
\end{figure}

Using the above fit parameters one can give expectations for the charm
structure function and $R$, the ratio of longitudinal to transverse
structure functions.  The uncertainty on the charm structure function
is gauged by varying the charm mass between $1.3$ and $1.7$~GeV. The
results are shown in figure~\ref{fig:f2c}. There is some indication that 
our curves are low compared to the data.

\begin{figure}
  \begin{center}
    \epsfig{file=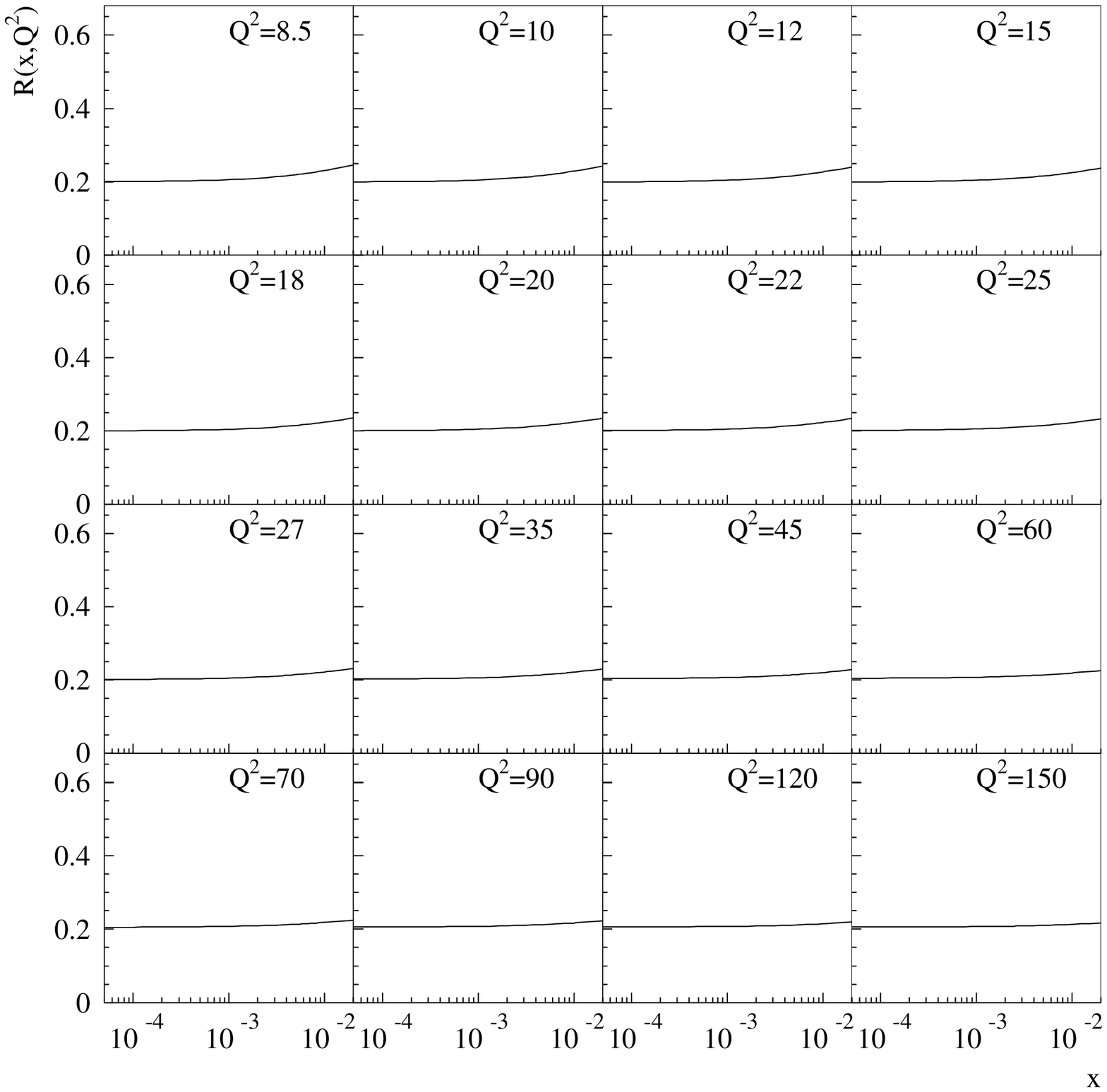,width=0.8\textwidth}
  \end{center}
  \caption{$R$}
  \label{fig:r}
\end{figure}

The results for $R$, shown in figure~\ref{fig:r}, are concentrated
around $R\simeq0.2$ fairly independently of $x$ and
$Q^2$. This seems to be fairly typical of results from BFKL-based
calculations (see for example \cite{NPRW}).

Before moving on to a discussion of final-state quantities, it is
perhaps worth commenting on some rather unexpected properties of the
$T=1-z$ form of the CCFM equation: as one decreases the cutoff $\mu$,
the steepness of the small-$x$ growth \emph{decreases}.  The reason
for this unusual behaviour, is that increasing $\mu$ removes from the
virtual corrections a region which was already dynamically suppressed
in the real emissions (see the discussion in section~\ref{sec:expn})
thus eliminating a negative contribution to the evolution, and
increasing the small-$x$ growth.  On the other hand, as is more
normal, increasing $\as^{(0)}$ increases the small-$x$ growth (though
not as substantially as in the pure BFKL case). As a result the best
fits tend to favour a diagonal band in the $\alpha_0$, $\ln \mu$
space.

This is to be contrasted with the behaviour that one sees with the
BFKL equation or with the CCFM equation with $T=1$, where the
small-$x$ growth increases as one decreases $\mu$ (there is none of
the dynamic non-cancellation between real and virtual parts which is
seen with $T=1-z$). We note that if one tries to fit the structure
function using BFKL, or CCFM with $T=1$ evolution, then to suppress
the small-$x$ growth sufficiently to fit the data, one requires an
unphysically large collinear cutoff $\mu\simeq3$~GeV.

\section{One-particle inclusive quantities}
\label{sec:1pi}

A number of final-state properties measured at HERA can be
approximated by one-particle-inclusive quantities. The latter are
relatively straightforward to calculate with the CCFM equation once
one has in place an approach for solving for the evolution of the
unintegrated gluon distribution.

Suppose that one is interested in determining the number density of
particles entering into a certain region of phase space with a given
transverse momentum $\qbt$ and rapidity $\ebt$.  It is convenient to
introduce an intermediate gluon density $\cB_{\qbt\ebt}(x,k,p)$ which
is obtained by considering configurations with any number of
emissions, followed by an emission into the region of interest:
\begin{multline}
  \label{eq:cB}
  \cB_{\qbt\ebt}(x,k,p) = 
  \int_x^1 \frac{dz}{z} \int \frac{d^2q}{\pi q^2} \,
  \asb\left(q_t\right) 
  \; \cA(x/z,|\vk+\vq_t\,|,q)\\ \times
  \De(z,k,q) \; \Theta(p-zq) \;\delta(\qbt-q_t)\;
  \delta\left(\ebt - \ln \frac{zq}{2xE_p} \right)\,.
\end{multline}
To obtain a full one-particle-inclusive density
$\cC_{\qbt\ebt}(x,k,p)$, one 
then allows any number of further emissions,
\begin{multline}
  \label{eq:cC}
  \cC_{\qbt\ebt}(x,k,p) = \cB_{\qbt\ebt}(x,k,p) \;+ \\
  \int_x^1 \frac{dz}{z} \int \frac{d^2q}{\pi q^2}\,
  \asb\left(q_t\right) 
  \; \cC_{\qbt\ebt}(x/z,|\vk+\vq_t\,|,q)
  \;\De(z,k,q) \;\Theta(p-zq)
  \,.
\end{multline}
Finally one performs a convolution with the boson-gluon fusion matrix
element, as in \eqref{eq:bgfconv}, to obtain the
single-particle-inclusive differential cross section:
\begin{equation}
  \label{eq:1pi}
  \frac{d\sigma}{dx\,dQ^2\,d\qbt\,d\ebt}\,=\,
   \int_x^1 \frac{dz}{z} \int \frac{dk}{k}
  {\hat\sigma} (z,k^2,Q^2,m^2_q) \cC_{\qbt\ebt}(x/z,k,p)\,,
\end{equation}
where in ${\hat\sigma}$ one sums over $F_2$ and $2xF_1$,
and over quark flavours. The contribution to the final state from the
quark box is not included.

\subsection{Transverse energy flows}
\label{sec:etflow}

\begin{figure}
  \begin{center}
    \epsfig{file=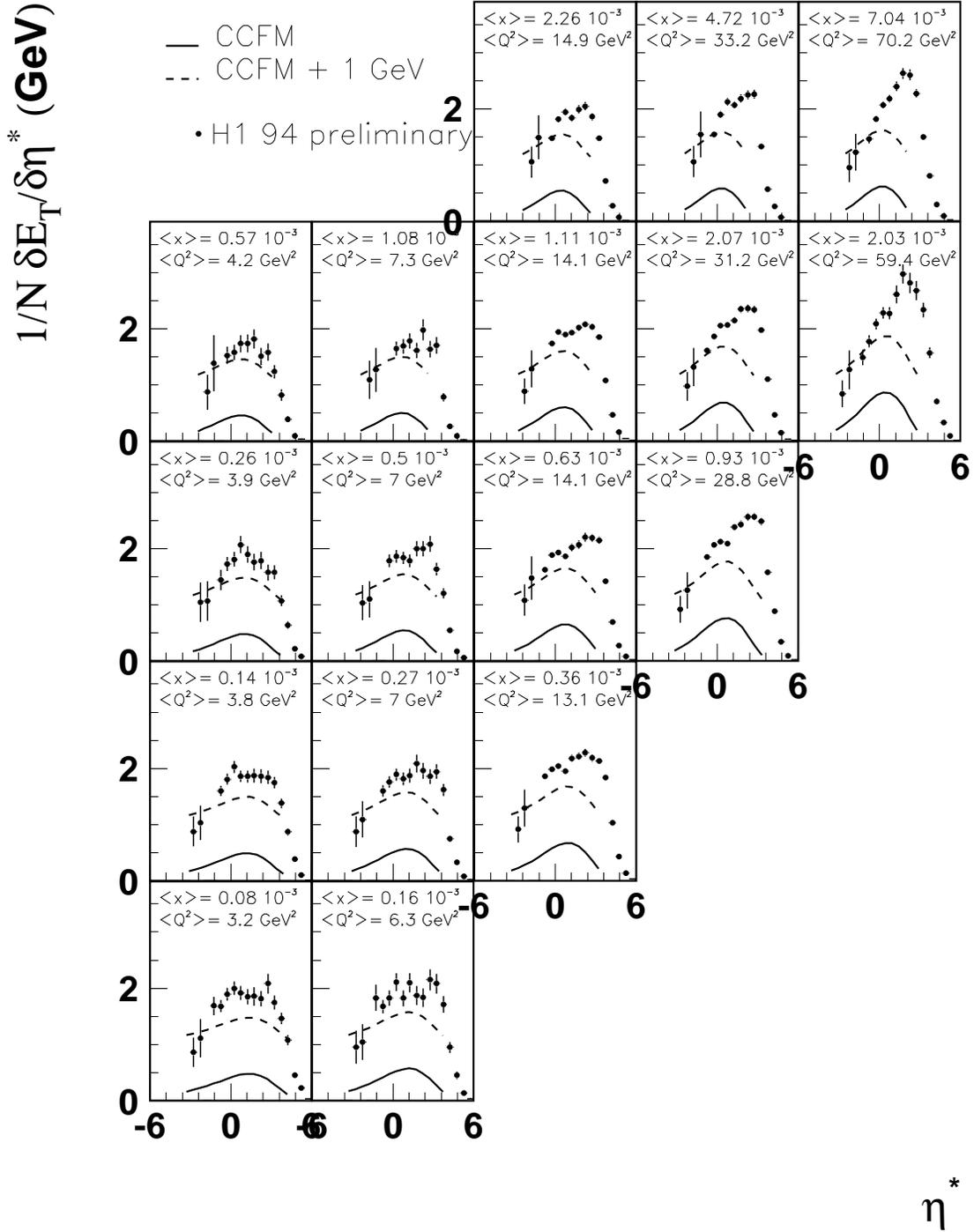, height=0.80\textheight}
    \caption{$E_t$ in various bins of $x$ and $Q^2$; A comparison is
      made with H1 1994 preliminary data \cite{H1Et}; the
      pseudorapidity is calculated in the photon-proton centre-of-mass
      system, with the proton direction being at negative rapidities.}
    \label{fig:etflow}
  \end{center}
\end{figure}

The mean transverse energy flow is given by
\begin{equation}
  \label{eq:etflow}
  \frac{dE_t}{d\eta}(x,Q^2) = \left(\frac{d\sigma}{dx\,dQ^2}\right)^{-1}
  \int dq_t \,q_t\,\frac{d\sigma}{dx\,dQ^2\,dq_t\,d\eta}\,.
\end{equation}
Results are compared with H1 data \cite{H1Et} in
figure~\ref{fig:etflow}. One sees that they are uniformally
low. There are probably two main reasons why this is so: firstly we
have neglected soft radiation from the $t$-channel gluon, namely the 
$$
\frac{dz}{1-z}
$$
part of the splitting function. It is responsible for the bulk of
the multiplicity at small transverse energies. As such it is a
formally subleading term. However given that ``small transverse
energies'' may mean of the order $1$~GeV, and that the actual
transverse energies that one is observing are about $2$~GeV, one can
immediately see that soft radiation may contribute significantly.
Another element comes from hadronisation, at a level analogous to a
$1/Q$ correction in $e^+e^-$ or DIS event shapes, from which one might
expect a contribution of the order of $0.5$~GeV per unit rapidity (though
this amount is tightly correlated with what one takes as the
perturbative contribution \cite{DW}).

In association with these ideas, it is interesting to note that if one
simply adds $1$~GeV of transverse energy per unit rapidity to the
curves obtained with \eqref{eq:etflow} then one finds a somewhat
better (but not perfect) agreement with the data. The agreement breaks
down in the highest rapidity bins at higher $Q^2$ values --- this is to
be expected given that we are not including the transverse energy that
comes from the quark box.

Using other parameter sets that are consistent with the structure
function data has little effect on the $E_t$ flow.

\subsection{Particle spectra}
\label{sec:ktspect}
\begin{figure}
  \begin{center}
    \epsfig{file=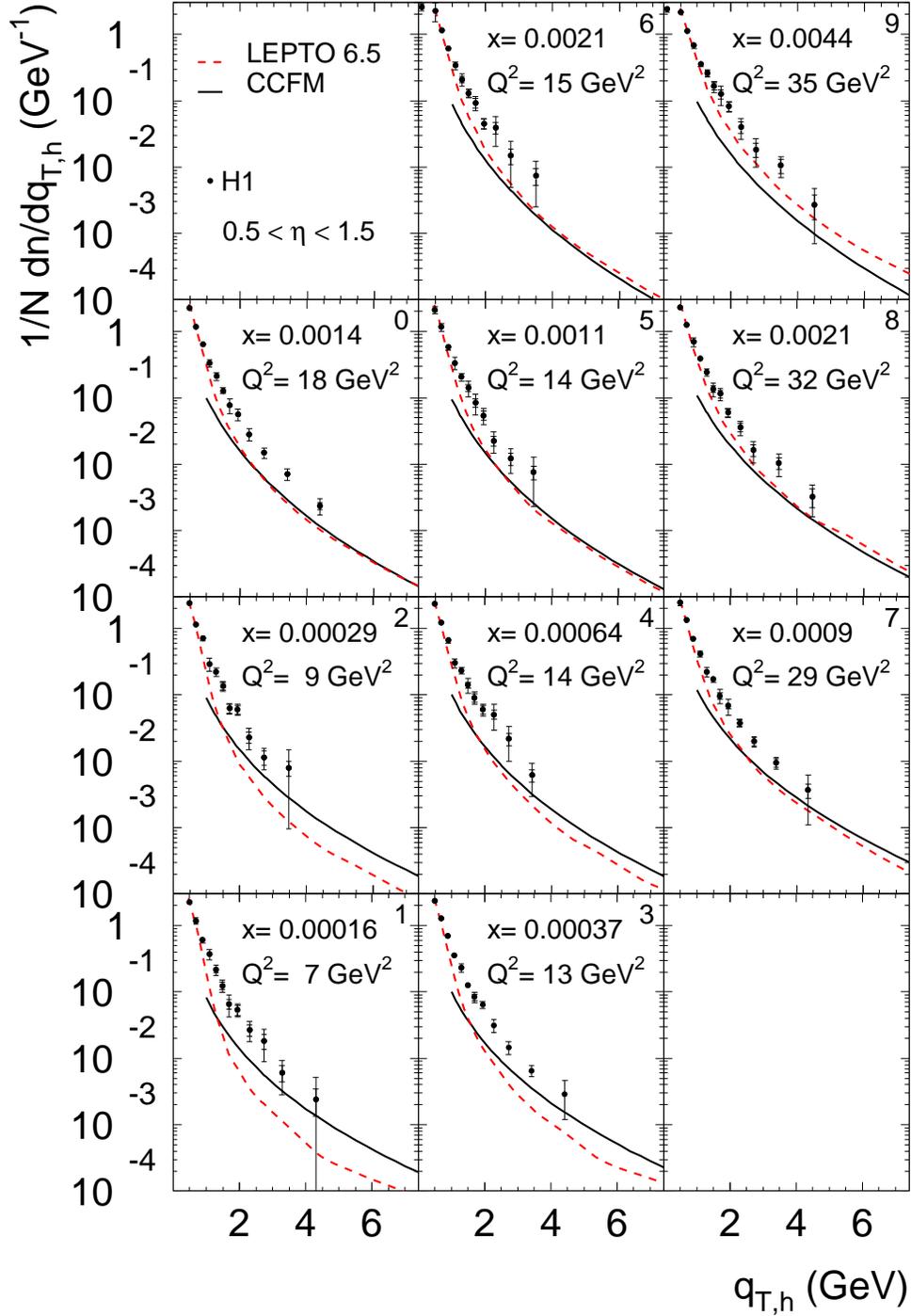,height=0.8\textheight}
    \caption{The charged hadron $k_t=\qth$ spectrum in various bins of
      $x$ and $Q^2$ for $0.5<\eta_{\mbox{\scriptsize cms}}<1.5$. The
      data are from the H1 collaboration \cite{H1kts}.}
    \label{fig:ktspect}
  \end{center}
\end{figure}

To avoid the problems associated with the $E_t$ flow, one should
consider a measurement which concentrates on high-transverse momenta,
and is consequently less sensitive to hadronisation and to the
mistreatment of the emission of soft particles. Such an observable is
the charged hadron transverse momentum ($q_{t,h}$) spectrum for a
given rapidity.  Since the one-particle inclusive cross section
calculated in \eqref{eq:1pi} is for emitted gluons, to allow a
comparison with the data it is necessary to perform the convolution
with the appropriate fragmentation function to obtain the spectrum for
hadrons. The spectrum for a hadron of type $h$ is
\begin{equation}
  \label{eq:ktspect}
    \frac{dn}{d\qth\,d\eta}(x,Q^2) = 
    \left(\frac{d\sigma}{dx\,dQ^2}\right)^{-1}
    \int_\qth \frac{d\qbt}{\qbt} \,D^h_g(\qth/\qbt,\qbt^2)\,
    \frac{d\sigma}{dx\,dQ^2\,d\qbt\,d\eta}\,.
\end{equation}
We make the approximation that the direction of the resulting hadron
is the same as that of the parent gluon and use the NLO fragmentation
functions of Binnewies \textit{et al.}, as given in \cite{frag}. At
our level of accuracy, it would have been equally valid to use the LO
fragmentation functions.

We compare with the preliminary results from the H1 1994 data
\cite{H1kts} in figure~\ref{fig:ktspect}. At low transverse momenta
the spectra that we obtain are much lower than the data --- this is as
expected given the lack of soft-radiation and hadronisation
corrections discussed in the previous section.

For the large-$q_{t,h}$ part of the spectrum there is moderate
agreement for a range of $x$ and $Q^2$ bins, though the CCFM results
seem systematically low. At larger $x$ this can be explained by the
fact that we are not counting the charged particles that come from the
fragmentation of the quark box. At lower $x$ one can however say that
the CCFM results are favoured compared to those from a DGLAP-based
Monte Carlo event generator, such as \lepto \cite{LEPTO}, though they
are still not perfect.

The sensitivity of these results to the evolution parameters is
moderate: altering $k_0$ by a factor of two (and the other
fit parameters appropriately) affects the normalisation by about
$25\%$.

\subsection{Forward jets}
\label{sec:fj}

Another quantity that is supposed to be particularly sensitive to
small-$x$ dynamics is the forward-jet cross section, originally
proposed in a slightly different form by Mueller and Navelet for
proton-proton collisions \cite{MuNa} and then adapted to DIS
\cite{MuFJ,DISFJ}.  One of the main difficulties in obtaining a reliable
prediction of this quantity is that experimentally one measures a jet,
using a jet algorithm. The best that one can do with a
one-particle-inclusive cross section is to associate with an emitted
gluon a jet with the same energy and direction. This neglects the
effect of multiple emission from the $t$-channel gluon, partonic
showering and hadronisation, all of which play a role which varies
according to the jet algorithm: as shown by the ZEUS collaboration
\cite{ZEUSfj}, the choice of jet-algorithm can quite easily affect the
results by about $15\%$. So in some sense this is the highest level of
precision with which one can compare to a single-particle-inclusive
approach with the data. Another element is that events in which 2 jets
satisfy the selection cuts are treated differently experimentally (one
counts only one of them), and one-particle-inclusively (one counts
both of them) -- however such events occur rarely ($2\%$ of the time
in the ZEUS sample), and so this difference in treatment is of little
practical importance.

In most forward-jet rate calculations, rather than using CCFM (or
BFKL) evolution before the emission of the forward jet, as in
\eqref{eq:cB}, one uses DGLAP evolution (i.e.\ rather than $\cB$
depending on $\cA$, it depends on the normal DGLAP structure
functions, $xg(x)$ and $xq(x)$). The motivations for using DGLAP
evolution are that at relevant values of $x_{jet}$ (which can be of
the order of $0.1$) one has a good knowledge of the DGLAP parton
densities; our CCFM-based densities in that region may not be
sufficiently constrained by the small-$x$ $F_2$ data, and in any case
do not include quarks. On the other hand, with CCFM evolution one has
the full $(k,p)$ dependence of the gluon density, whereas with DGLAP
evolution this information must be added in. 

To use the normal DGLAP structure functions before the emission, one
follows exactly the same procedure as before, except that one replaces
\eqref{eq:cB} with
\begin{multline}
  \label{eq:cBdglap}
  \cB^{(\mathrm{DGLAP})}_{\qbt\ebt} (x,k,p) = 
  \int_x^1 \frac{dz}{z} \int \frac{d^2q}{\pi q^2} \,
  \asb\left(q_t\right) 
  \; \frac{x}{z} F\left(\frac{x}{z}, q_t^2\right) \delta^2(\vk+\vqt)
  \\ \times \Delta(z,k,q)\;
  \Theta(p-zq) \;\delta(\qbt-q_t)\;
  \delta\left(\ebt - \ln \frac{zq}{2xE_p} \right)\,,
\end{multline}
with $\vq_t = (1-z)\vq$ and 
\begin{equation}
F(x,q_t^2) = \left[ g\left({x},q_t^2\right)
    + \frac{\cf}{\ca}
    \left(q(x,q_t^2)+ {\bar q}(x,q_t^2) \right)\right],
\end{equation}
where $g$ and $q$ are usual DGLAP parton densities in the proton (and
the sum over flavours is implicit for the quark and antiquark
densities). Fixing $k$ through $\delta^2(\vk+\vqt)$ corresponds to the
approximation of strong transverse-momentum ordering (which should be
valid in the region of $x_{jet}$ of interest).

The cuts used for the forward-jet measurement tend to be considerably
more complicated than for the energy-flow or particle spectra. As a
result the simplest way to obtain a prediction is to use a Monte Carlo
integration method which generates $x,Q^2$ values, a four vector for
the forward gluon (or quark), and a corresponding weight. To these one
applies the experimental cuts as if the forward parton were a jet, and
finally adds the weights of the configurations that survive.

\begin{figure}
  \begin{center}
    \epsfig{file=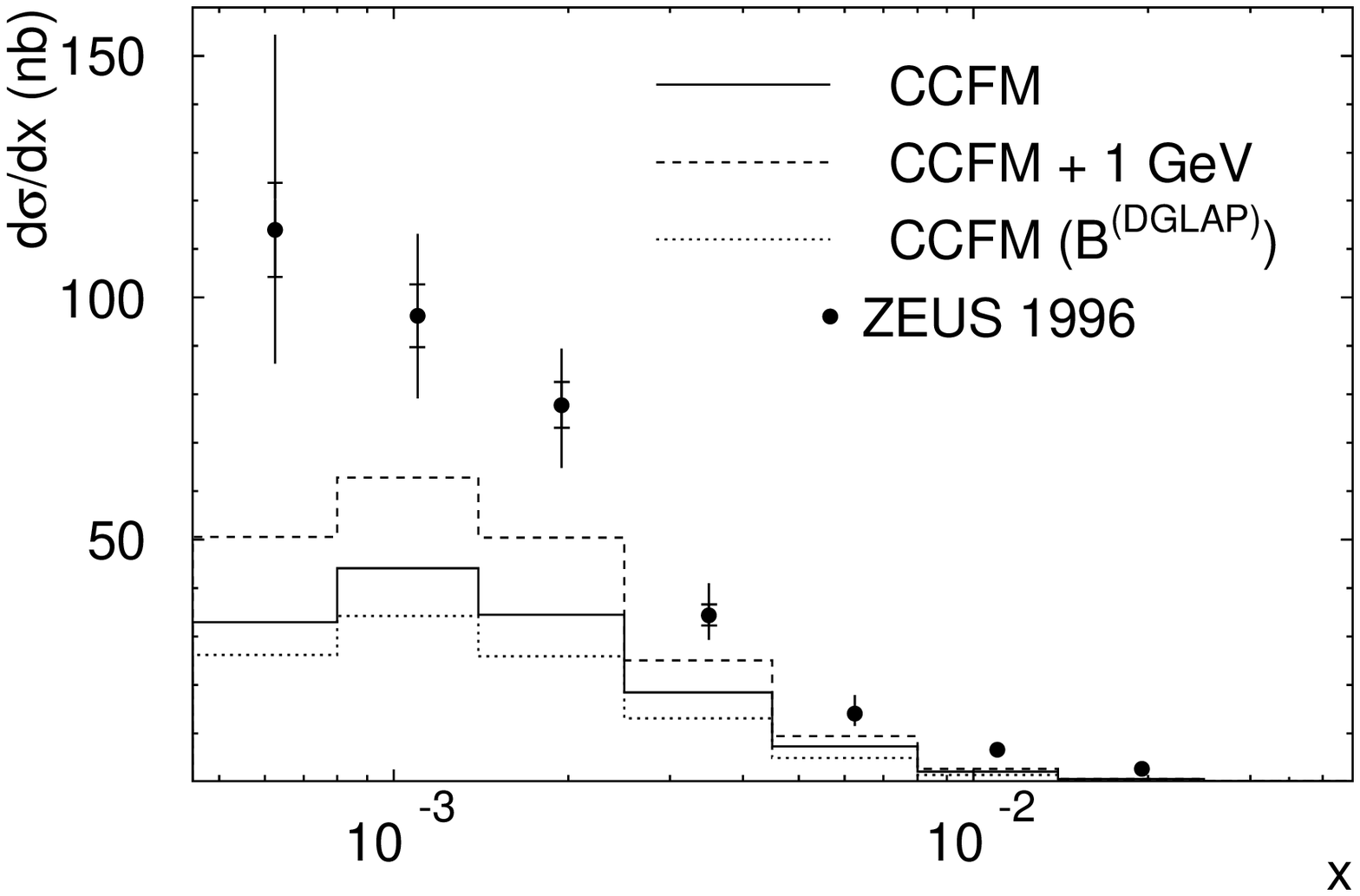, width=0.8\textwidth}\\
    \epsfig{file=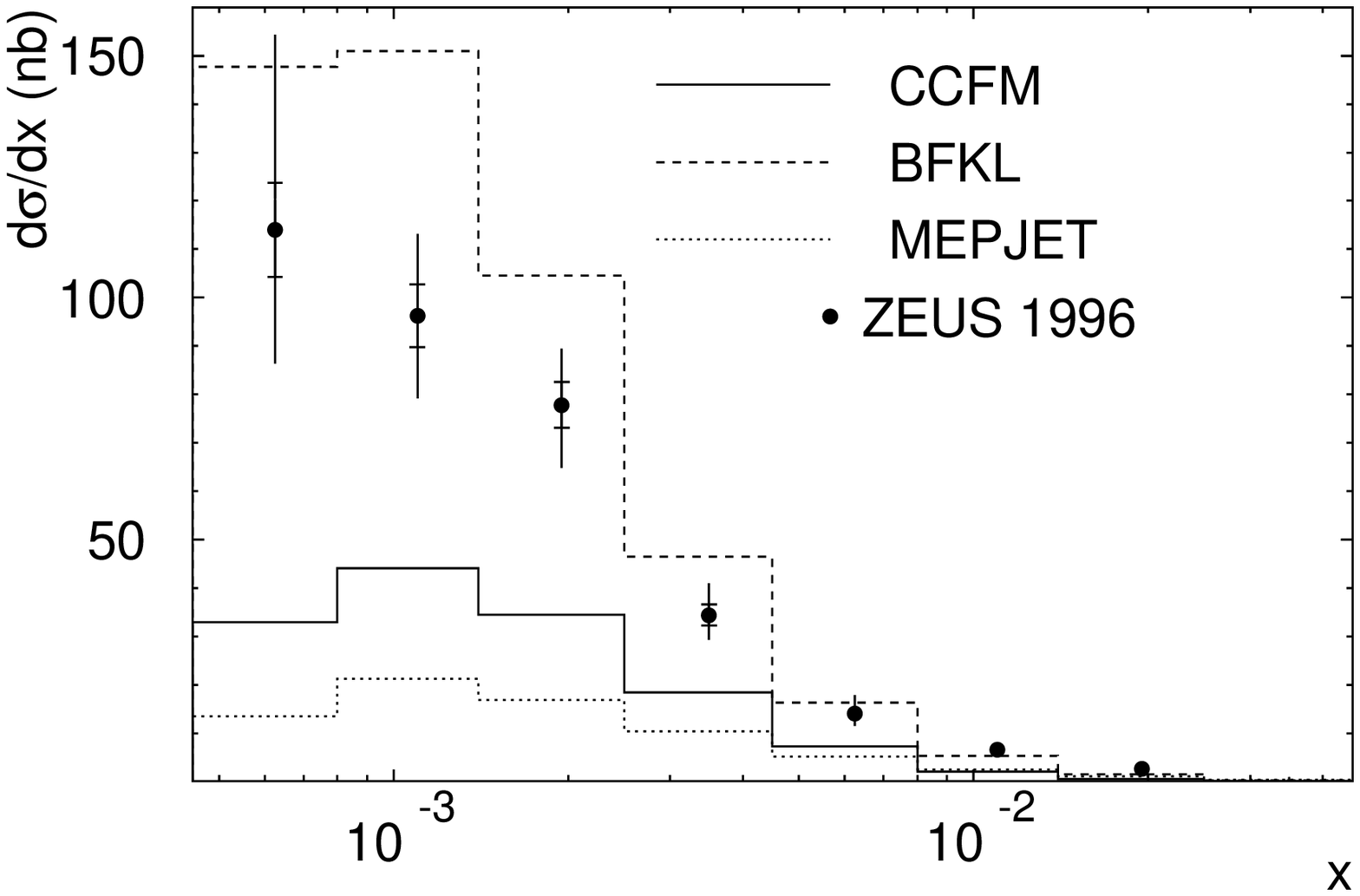, width=0.8\textwidth}
    \caption{The forward-jet cross section, and the results of 
      the ZEUS measurement \cite{ZEUSfj}. The upper figure shows three 
      different approaches to calculating the results using the CCFM
      equation, while the lower figure compares the CCFM results with
      other theoretical predictions \cite{BW,MEPJET}. }
    \label{fig:fj}
  \end{center}
\end{figure}

We will compare our calculations with the ZEUS measurement
\cite{ZEUSfj}. The cuts are: $0.00045<x<0.045$, $\etj>5$~GeV,
$0.5<\etj^2/Q^2<2$, the energy of the scattered electron
$E_{e'}>10$~GeV, $y_{Bj}>0.1$, $\theta_{Jet}<8.5$~degrees,
$x_{Jet}>0.036$, and the jet must be in the target hemisphere of the
Breit frame.

Figure \ref{fig:fj} shows the ZEUS results and a calculation which
includes CCFM evolution both before and after the forward-jet
emission. The CCFM cross section is consistently too low everywhere.
In the higher-$x$ bins this is not unexpected since there are other
mechanisms of importance there which are not being considered (e.g.\ 
the forward jet can come from one of the legs of the quark box). But
at low $x$ it is more difficult to justify. To examine the degree of
uncertainty on our result, we have tried to simulate the effect of
hadronisation and ``mis-reconstruction'' from a jet algorithm by
adding 1~GeV of transverse energy to all jets, while maintaining their
rapidity. This has a greater effect at small $x$ than at larger $x$,
due to the fact that the cut on $\etj^2/Q^2$, in conjunction with the
correlation between $x$ and $Q^2$ (the cut on $y_{Bj}$) causes jets in
events with larger $x$ to have a larger transverse momentum and hence
be less sensitive to hadronisation. The effect of this is not
inconsistent with the $\cO{15\%}$ jet-algorithm dependence mentioned
above. In any case it is not sufficient to bring the CCFM curves into
agreement with the data. The use of the DGLAP distribution for the
incoming partons, rather than the CCFM distribution, tends to lower
the results --- so that does not help either. Changing the parameter
sets for the evolution also had little effect, affecting only
moderately ($\sim 20\%$) the normalisation. We note that a similar
calculation carried out with the cuts used by H1 also falls below their
recent data \cite{H1fj96}.

For reference we include also a comparison of our results with those
of \cite{BW}, where a normal BFKL-based calculation of the forward-jet
cross section was carried out: the data are considerably closer to the
results of \cite{BW} than to ours.  We have also included the results
of a NLO calculation using MEPJET \cite{MEPJET}, as presented in the
ZEUS forward-jet paper \cite{ZEUSfj}. For this kind of observable NLO
is the first order which contributes, so the fact that the CCFM-based
results are above the MEPJET prediction is to be expected.

\begin{figure}
  \begin{center}
    \epsfig{file=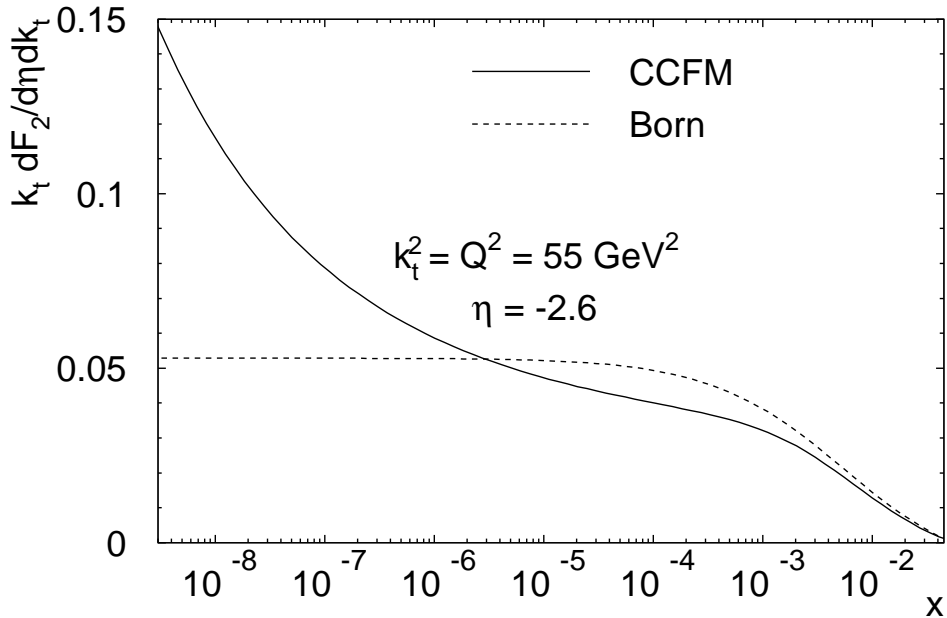, width=0.8\textwidth}\\
    \caption{A ``toy'' forward jet measurement, comparing the result 
      with evolution to that without evolution (Born). In both cases
      the input parton distribution is the DGLAP one.} 
    \label{fig:fjfq}
  \end{center}
\end{figure}

Why does the version of the CCFM equation that is being used here lead
to forward-jet results which are not satisfactory? The forward-jet
measurement that is used experimentally does not cast much light on
this, in part because of the numerous cuts, which lead to correlations
between $x$ and $\etj$ and restrict the available $x$ range. So it is
instructive to examine a forward-jet-like (where the forward-jet
cannot come from the quark box) measurement which is free of these
problems, such as
\begin{equation}
  \label{eq:fjtoy}
  \etj \,\frac{dF_2(x,Q^2)}{d\eta\, d\etj}\,,
\end{equation}
as a function of $x$, for fixed $\etj$, $Q^2$ and $\eta$. At very
small $x$ we know that in the absence of small-$x$ evolution (the Born
approximation) it should go to a constant, independent of $x$. So what
we can do is compare the CCFM result for this quantity with a similar
quantity worked out in the Born approximation, namely using
\begin{multline}
C^{(BORN)}_{\qbt\ebt}(x,k,p)=
  \int_x^1 \frac{dz}{z} \int \frac{d^2q}{\pi q^2} \,
  \asb\left(q_t\right) 
  \; \frac{x}{z} F\left(\frac{x}{z}, q_t^2\right) \delta^2(\vk+\vqt)
  \\ \times 
  \Theta(p-zq) \;\delta(\qbt-q_t)\;
  \delta\left(\ebt - \ln \frac{zq}{2xE_p} \right)\,.
\end{multline}
The result is shown in figure~\ref{fig:fjfq}.

Concentrating first on the Born result, we see that it takes a
considerable range in $x$ to reach its constant value (bearing in mind
that $x_{jet} \simeq 0.06$): assuming that the incoming gluon has
transverse momentum $Q$, the kinematics of the quark box require the
momentum fraction of the incoming gluon, $x_g$, to satisfy $x_g>
2x_{Bj}$ --- to obtain the full asymptotic cross section $x_g$ should
be larger than that limit by an order of magnitude (the cross section
goes as $dx_g / x_g^2$).  But $x_g$ itself must satisfy $x_g \ll
x_{jet}$ otherwise the cross section is sensitive to the DGLAP parton
distributions at $x$ values larger than $x_{jet}$, and the gluon
distribution drops quite fast with increasing $x$.

What is surprising is that the full CCFM result should drop
\emph{below} the Born result for such a considerable range in $x$, and
only for $x<10^{-5}$ does it start to overtake the Born result: the
CCFM evolution takes a long time to get started. This again is
probably a consequence of the large virtual corrections, which can be
compensated only when there is a sufficiently asymptotic gluon
transverse-momentum distribution --- and that which arises immediately
after the forward-jet emission is far from asymptotic.

\section{Conclusions}
\label{sec:conc}

The main aim of this paper has been to apply the angular ordered
small-$x$ formalism, in the form of the CCFM equation, to the
description of HERA data, with the purpose of determining whether, as
it stands currently, it is suitable as a basis for quantitative
phenomenology.

As a first step it was necessary to decide on the treatment of $1-z$
kinematical factors in the virtual corrections ($T=1$ or $T=1-z$, in
\eqref{eq:nsff}). The final choice was made on the basis that $T=1-z$
gives an asymptotic exponent which is considerably lower than $T=1$
and LL BFKL, and roughly in accord with phenomenological expectations
and with first estimates from a partial resummation of NLL corrections
to the BFKL equation. This choice does however have the property that
its NLL corrections are four times larger than the true NLL
corrections to the BFKL equation. The part of the splitting function
which is finite as $z\to0$ was left out, partly because it is not
clear how it should be included, partly because whichever method of
inclusion one chooses, there are considerable technical difficulties
in its implementation.  We will return to the discussion of these
points shortly.  To constrain the infra-red properties of the
evolution, we used the $F_2$ structure function; we then examined some
of the classic ``BFKL'' structure-function and final-state signals:
$F_2^c$, $R$, the transverse energy flow, the charged hadron
transverse momentum spectrum and the forward-jet cross section.

The transverse energy flow is much lower than the data, but this was
to be expected: both the $z\to0$ finite part of the splitting function
and non-perturbative contributions are expected to give large
(though formally subleading) contributions.

The failure to reproduce the forward-jet cross section data is less
acceptable. An analysis of the problem indicates that after the
emission of the forward jet, the small-$x$ evolution has a long period
in which there is no growth of the cross section --- non-asymptotic
effects are important. The exact reason for this may well be connected
with the non-trivial interplay between real and virtual contributions,
which leads also to the large NLL corrections to the evolution. The
importance of non-asymptotic effects may also be responsible for the
lowness of the predictions for $F_2^c$ and the charged-particle
transverse momentum spectra (though the results for the latter are
better than those from DGLAP-based approaches, at least for some
values of $x$ and $Q^2$).


It is interesting to note that the LDC model \cite{LDCorig,LDCmc}
displays problems similar to those found here. The physics contained
is somewhat different (e.g.\ the evolution is symmetric) but the
general approach is somewhat similar --- they too use a fit to the
$F_2$ structure function as a constraint on the evolution parameters.
The fit to the structure function data works well, but again a number
of final state quantities are too low.  It is not clear whether the
problems in the LDC model are of the same nature as those in our CCFM
implementation (i.e.\ due to non-asymptotic effects), but it might be
a matter worthy of further investigation.

As a consequence of the problems described here, one is essentially
led to rule out the CCFM equation with $T=1-z$ and without the $z\to0$
finite part of the splitting function. It is tempting to suggest that
we should have used $T=1$ and then included the $z\to0$ finite part of
the splitting function (in some arbitrary way).  This would be the
physics of \SX\ program --- which so far has only been used for the
study of final states in which the photon-gluon fusion produces charm
quarks \cite{Biernat}. Since most final state studies at HERA do not
select for the presence of open charm, there is relatively little data
for comparison, so it would be worthwhile extending that study to the
case with the full mix of quarks in the photon-gluon fusion.

In the longer term we need to work towards a theoretical approach free
of the current ambiguities.  Much of the basic information is probably
already at our disposal, in the form of the NLL corrections to the
BFKL kernel \cite{NLLBFKL,CCFL}. We should be looking for an evolution
equation which rather just than giving a ``reasonable exponent'' (a
condition satisfied both by the LDC and by the form of the CCFM
equation described in this paper), explicitly reproduces the structure
of the NLL corrections (whether the part associated with
symmetrisation, of with the $z\to0$ finite part of the splitting
function), and which resums these corrections in a sensible way. Of
course it should also satisfy basic properties associated with
coherence in the final state.  Another important element lies in the
calculation of the boson-gluon fusion hard matrix element to
next-to-leading order. After the completion of such a programme we
should be in a relatively strong position to carry out quantitatively
meaningful BFKL phenomenology. It may take a while for such a stage
to be reached --- but in the mean time it is of paramount importance
that experimental measurements of BFKL signals continue.

\acknowledgments 

We would like to thank J. Bartels, N. Brook, T. Carli, M. Ciafaloni,
Yu.L.  Dokshitzer, K. Golec-Biernat, G. Gustafson, M. Kuhlen, L.
L\"onnblad, M. Ryskin, M. Riveline, B.R. Webber and M. W{\"u}sthoff
for helpful discussions.

\appendix

\section{Photon-gluon fusion} 
\label{sec:bgf}

In this appendix we outline the calculation of the one-loop
photon-gluon fusion hard matrix element.  The results presented have
been derived with the aid of the HIP package for
Mathematica \cite{MathHIP}.

\begin{figure}
  \begin{center}
    \epsfig{file=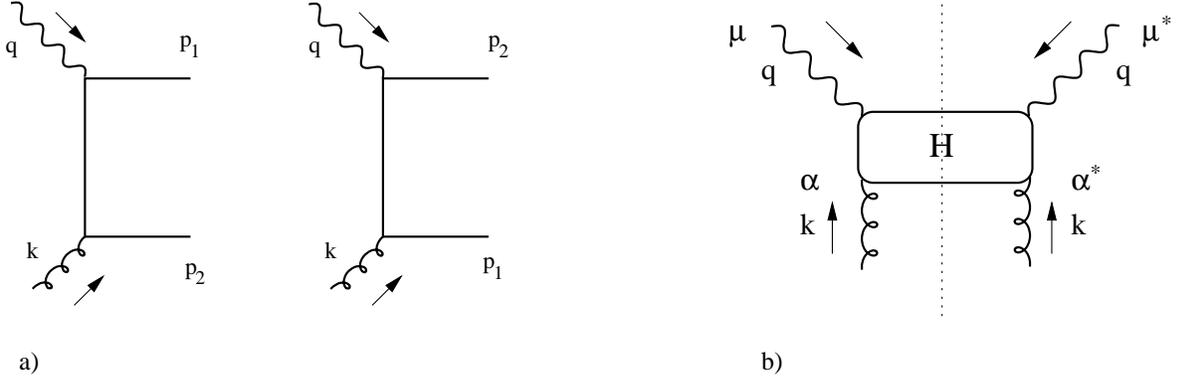, width=\textwidth}
    \caption[Hard matrix element diagrams]{ {\bf a)} The one loop diagrams
      contributing to photon gluon fusion vertex. {\bf b)} The squared 
      amplitude Lorentz indices.}
  \label{fig:diagra}
  \end{center}
\end{figure}

The starting point is the tree-level amplitude, coming from the
diagrams illustrated in figure~\ref{fig:diagra}a:
\begin{eqnarray}
  \label{eq:Mdef}
  {\cal M}^{\mu \al} &=& -i e_Q g_s \,t^a \; \bar{u} (p_1)
  \left[ { \gamma^\mu \; [(p_1-q)\cdot\gamma +m]\; \gamma^\al \over
      t-m^2} + \right. \\ \nonumber
  &&\left.
    { \gamma^\al \; [(p_1-k)\cdot\gamma +m]\; \gamma^\mu \over u-m^2 }
  \right] v(p_2) \;.
\end{eqnarray}
Here $e_Q$ is the fractional quark charge, $g_s$ is the strong
coupling constant, $t^a$ are the SU(3) colour generators, $m$ is the
quark mass, $p_1$ and $p_2$ are the outgoing quark and anti-quark
momenta and $q$ and $k$ denote respectively the photon and gluon
momenta\footnote{In this appendix $k$ is the gluon four-momentum, and
  $\kt$ the transverse momentum, whereas in the rest of the paper we
  denote by $k$ the latter quantity.}.  The Mandelstam variables are
defined as $s=(k+q)^2$, $t=(q-p_1)$ and $u=(q-p_2)^2$.

The hard matrix tensor $H^{\mu \mus, \al \als}$ is obtained by
squaring the amplitude, averaging over the colour of the incoming
gluon, summing over the final state (i.e. quark and anti-quark)
degrees of freedom and integrating over the two-particle phase space
$d\phi\,(p_1,p_2)$:
\begin{equation}
  \label{eq:Hdef}
  H^{\mu \mus, \al \als} (q,k) = \int d\phi\,(p_1,p_2) \;
  \overline{\sum_{g}} \; \sum_{q ,\bar{q}} 
  {{\cal M}^{\mu \al}}^\dagger \! {\cal M}^{\mus \als} \, .
\end{equation}
It depends only on the photon and gluon Lorentz indices and momenta
(figure~\ref{fig:diagra}).

The ``reduced cross section'' ${\hat\sigma}_{F_2}$ appearing in
\eqref{eq:bgfconv} is then defined through
\begin{equation}
  \label{eq:sigdef}
{\hat\sigma}_{F_i} =
  \frac{\kt^2}{2 \xf}\;
  \hat{w}_{i, \mu \mus} \; H^{\mu \mus \al \als}(q,k) \;
  \hat{P}_{\bfkl, \al \als} \; ,
\end{equation}
where the photon Lorentz indices are contracted over the usual
projector used in DIS to extract the structure function $F_2$
and the longitudinal structure function $F_L$:
\begin{eqnarray}
  \label{eq:w2def}
  w_2^{\mu \mus} &=&
  {1\over p \cdot q}
  \left(p^\mu - {p\cdot q \over q^2} q^\mu \right)
  \left(p^{\mus} - {p\cdot q \over q^2} q^{\mus} \right)
\\
  w_L^{\mu \mus} &=&
  {1\over p \cdot q} \left[
   q^2 \left( g^{\mu \mus} - {q^\mu q^{\mus} \over q^2}\right) -
  \left(p^\mu - {p\cdot q \over q^2} q^\mu \right)
  \left(p^{\mus} - {p\cdot q \over q^2} q^{\mus} \right) \right]
\end{eqnarray}
and the gluon Lorentz indices are contracted over the projector
\begin{equation}
  \label{eq:BFKLproj}
  \hat{P}_{\bfkl}^{\al \als} = 
  \left. \frac{k_{\perp}^\al k_{\perp}^{\als}}{-k^2_{\perp}} 
  \right|_{k = \xf \, p + k_{\perp}} =
  \frac{\xf^2}{\kt^2} p^{\al} p^{\als} \; ,
\end{equation}
coming from the high energy (or $\kt$) factorisation prescription
\cite{CCH,CollinsEllis,LevinEtAl}.  Additionally the prescription
specifies that in \eqref{eq:sigdef}, the incoming gluon momentum
should be approximated by $k = \xf p + k_\perp$ with $p$ the hadron
momentum, so that $k\cdot p = 0$ and $k^2 = k_\perp^2 = - \kt^2$.

After the contractions with the two projectors, the reduced cross
section acquires a dependence on the proton momentum (i.e.\ it is not
just a function of the photon and gluon momenta), which complicates
the phase-space integration for the outgoing quarks. As a result it
seems very difficult to carry out this integration analytically, and
the answer is usually left in the form of a Mellin transform, or of a
double integral which must be evaluated numerically
\cite{CCH,CollinsEllis,LevinEtAl,BW,KLM}.

In this appendix we show that it is possible to carry out the phase
space integration analytically, but this must be done before the
contraction with the projectors. Accordingly, we expand $H^{\mu \mus,
  \al \als}(q,k)$ over a tensor basis $\{T_i^{\mu \mus, \al
  \als}(q,k)\}$, taking into account the symmetry of $H^{\mu \mus, \al
  \als}$ with respect to both the $\mu,\mus$ and $\al,\als$ pair of
indices, and its transversality with respect to $q$ and $k$ (i.e.
$q_\mu H^{\mu \mus, \al \als}=0$, $k_\al H^{\mu \mus, \al \als}=0$).

First of all, let us introduce a four-dimensional basis $h_i^{\mu
  \nu}(a,b)$ for a two-index symmetric tensor which depends on two
four-momenta $a$ and $b$:
\begin{eqnarray}
 h_1^{\mu \nu}(a,b) &=& 
    - \left( g^{\mu \nu} - {a^\mu a^\nu \over a^2}\right) \,,\nonumber \\
 h_2^{\mu \nu}(a,b) &=&
    -{a^2\over \sqrt{(a\cdot b)^2 - a^2 b^2 }}
     \left(b^\mu - {a\cdot b \over a^2} a^\mu \right)
     \left(b^\nu - {a\cdot b \over a^2} a^\nu \right)\,, \nonumber \\
 h_3^{\mu \nu}(a,b) &=& {a^\mu a^\nu \over a^2}\,, \\
 h_4^{\mu \nu}(a,b) &=& {a^\mu b^\nu + b^\mu a^\nu \over a\cdot b} \,.\nonumber
\end{eqnarray}
We then construct a basis $B_i^{\mu \mus, \al \als}(q,k)$ for a
four-index tensor which depends on the two momenta $q$ and $k$ and
which is symmetric with respect to $\mu,\mus$ and to $\al,\als$.
There are 21 such independent structures.  Sixteen can be chosen as
the products
\begin{equation}
  B_{4(i-1)+j}^{\mu \mus, \al \als}(q,k) = h_{i}^{\mu \mus}(q,k)
  \; h_{j}^{\al \als}(k,q)\,,
\ \ \ \ i,j=1,\ldots,4\,,
\end{equation}
while for the others we take
\begin{eqnarray}
  B_{17}^{\mu \mus, \al \als}(q,k) &=&
\left( g^{\mu \al} g^{\mus \als} +  g^{\mu \als} g^{\mus \al}  \right)
\,, \nonumber \\
  B_{18}^{\mu \mus, \al \als}(q,k) &=&
   \frac{1}{q^2} \left(  
   g^{\mu \al} q^\mus q^\als +  g^{\mu \als} q^\mus q^\al +  
   g^{\mus \al} q^\mu q^\als +  g^{\mus \als} q^\mu q^\al \right)
\,, \nonumber \\
  B_{19}^{\mu \mus, \al \als}(q,k) &=&
   \frac{1}{k^2} \left(  
   g^{\mu \al} k^\mus k^\als +  g^{\mu \als} k^\mus k^\al +  
   g^{\mus \al} k^\mu k^\als +  g^{\mus \als} k^\mu k^\al \right)
\,, \\
  B_{20}^{\mu \mus, \al \als}(q,k) &=&
   \frac{1}{q\cdot k} \left(  
   g^{\mu \al} q^\mus k^\als +  g^{\mu \als} q^\mus k^\al +  
   g^{\mus \al} q^\mu k^\als +  g^{\mus \als} q^\mu k^\al \right)
\,, \nonumber \\
  B_{21}^{\mu \mus, \al \als}(q,k) &=&
   \frac{1}{q\cdot k} \left(  
   g^{\mu \al} k^\mus q^\als +  g^{\mu \als} k^\mus q^\al +  
   g^{\mus \al} k^\mu q^\als +  g^{\mus \als} k^\mu q^\al \right)
\,, \nonumber
\end{eqnarray}
Since $H^{\mu \mus, \al \als}(q,k)$ is explicitly transverse with
respect to both $q$ and $k$, let us now concentrate on the set of
symmetric tensors $T_i^{\mu \mus, \al \als} = \sum_{j=1}^{21} C_{ij}
B_j^{\mu \mus, \al \als}$ that satisfy $q_\mu T_i^{\mu \mus, \al
  \als}=0$ and $k_\al T_i^{\mu \mus, \al \als}=0$. There are 6 such
independent tensors, specified by
\renewcommand{\arraystretch}{1.3}
\begin{equation}
C^T = \left(
  \begin{array}{cccccc}
    {3\over 2} - {1\over 4 \rho} & {1\over 4 \rho} & 
    {1\over 4 \rho} & {{-1}\over 4 \rho} & {1\over {2 \rho }} & 
    {{-1}\over 4 \rho} \\ 
    {1\over 2} - {1\over 4 \rho} & 
    1 + {1\over 4 \rho} & {1\over 4 \rho} & {{-1}\over 4 \rho} & 
    {1\over {2 \rho }} & {{-1}\over 4 \rho} \\ 
    0 & 0 & 0 & 0 & 0 & 0 \\
    0 & 0 & 0 & 0 & 0 & 0 \\ 
    {1\over 2} - {1\over 4 \rho} & {1\over 4 \rho} & 
    1 + {1\over 4 \rho} & {{-1}\over 4 \rho} & {1\over {2 \rho }} & 
    {{-1}\over 4 \rho} \\ {1\over 2} + {1\over {2 \rho^2 }} + 
     {3\over 4 \rho} & {{-\left( 2 + 7 \rho  \right) }\over 
      {4 \rho^2 }} & {{-\left( 2 + 7 \rho  \right) }\over 
      {4 \rho^2 }} & 1 + {1\over {2 \rho^2 }} + {7\over 4 \rho} & 
    {{-\left( 2 + 3 \rho  \right) }\over {2 \rho^2 }} & 
    {{2 - \rho }\over {4 \rho^2 }} \\
    -1 + {1\over {2 \rho^2 }} + {3\over {2 \rho }} & 
    {{-\left( 1 + 5 \rho  \right) }\over {2 \rho^2 }} & 
    {{-\left( 1 + 5 \rho  \right) }\over {2 \rho^2 }} & 
    {{1 + 5 \rho }\over {2 \rho^2 }} & 
    {{-1 - 3 \rho }\over {\rho^2 }} & 
    {{1 + \rho }\over {2 \rho^2 }} \\
    {{-1}\over {2 \rho^2 }} & 
    {1\over {2 \rho^2 }} + {1\over {\rho }} & 
    {1\over {2 \rho^2 }} + {1\over {\rho }} & 
    {{-1}\over {2 \rho^2 }} - {1\over {\rho }} & 
    {{1 + \rho }\over {\rho^2 }} & {{-1}\over {2 \rho^2 }} \\
    0 & 0 & 0 & 0 & 0 & 0 \\ 
    -1 + {1\over {2 \rho^2 }} + {3\over {2 \rho }} & 
    {{-\left( 1 + 5 \rho  \right) }\over {2 \rho^2 }} & 
    {{-\left( 1 + 5 \rho  \right) }\over {2 \rho^2 }} & 
    {{1 + 5 \rho }\over {2 \rho^2 }} & 
    {{-1 - 3 \rho }\over {\rho^2 }} & 
    {{1 + \rho }\over {2 \rho^2 }} \\
    {{-1 - 3 \rho  + 2 \rho^2 }\over {2 (\rho - 1)  \rho^2 }} & 
    {{1 + 5 \rho }\over {2 (\rho - 1)  \rho^2 }} & 
    {{1 + 5 \rho }\over {2 (\rho - 1)  \rho^2 }} & 
    {{1 + 5 \rho }\over {2 (1 - \rho)  \rho^2 }} & 
    {{1 + 3 \rho }\over {(\rho - 1)  \rho^2 }} & 
    {{1 + \rho }\over {2 (1 - \rho)  \rho^2 }} \\
    {1\over {2 (\rho - 1)  \rho^2 }} & 
    {{1 + 2 \rho }\over {2 (1 - \rho)  \rho^2 }} & 
    {{1 + 2 \rho }\over {2 (1 - \rho)  \rho^2 }} & 
    {{1 + 2 \rho }\over {2 (\rho - 1)  \rho^2 }} & 
    {{1 + \rho }\over {\rho^2  - {{\rho }^3}}} & 
    {1\over {2 (\rho - 1)  \rho^2 }} \\
    0 & 0 & 0 & 0 & 0 & 0 \\ 
    {{-1}\over {2 \rho^2 }} & 
    {1\over {2 \rho^2 }} + {1\over {\rho }} & 
    {1\over {2 \rho^2 }} + {1\over {\rho }} & 
    {{-1}\over {2 \rho^2 }} - {1\over {\rho }} & 
    {{1 + \rho }\over {\rho^2 }} & {{-1}\over {2 \rho^2 }} \\
    {1\over {2 (\rho - 1)  \rho^2 }} & 
    {{1 + 2 \rho }\over {2 (1 - \rho)  \rho^2 }} & 
    {{1 + 2 \rho }\over {2 (1 - \rho)  \rho^2 }} & 
    {{1 + 2 \rho }\over {2 (\rho - 1)  \rho^2 }} & 
    {{1 + \rho }\over {\rho^2  - {{\rho }^3}}} & 
    {1\over {2 (\rho - 1)  \rho^2 }} \\
    {{-2 + 3 \rho  - 2 \rho^2 }\over 
      {4 (\rho - 1)  \rho^2 }} & 
    {{2 + \rho }\over {4 (\rho - 1)  \rho^2 }} & 
    {{2 + \rho }\over {4 (\rho - 1)  \rho^2 }} & 
    {{2 + \rho }\over {4 (1 - \rho)  \rho^2 }} & 
    {1\over {(\rho - 1)  \rho^2 }} & 
    {{-2 + \rho }\over {4 (\rho - 1)  \rho^2 }} \\
    -{1\over 2} + {1\over 4 \rho} & {{-1}\over 4 \rho} & 
    {{-1}\over 4 \rho} & {1\over 4 \rho} & {{-1}\over {2 \rho }} & 
    {1\over 4 \rho} \\ 
    {{-3 + 2 \rho }\over {4 (\rho - 1) }} & 
    {3\over {4 (\rho - 1) }} & 
    {3\over {4 (\rho - 1) }} & {3\over {4 - 4 \rho }} & 
    {1\over {-1 + \rho }} & {1\over {4 - 4 \rho }} \\
    {{-3 + 2 \rho }\over {4 (\rho - 1) }} & 
    {3\over {4 (\rho - 1) }} & 
    {3\over {4 (\rho - 1) }} & {3\over {4 - 4 \rho }} & 
    {1\over {-1 + \rho }} & {1\over {4 - 4 \rho }} \\
    {{-3 + 2 \rho }\over {4 \rho  - 4 \rho^2 }} & 
    {3\over {4 \rho  - 4 \rho^2 }} & {3\over {4 \rho  - 4 \rho^2 }} & 
    {3\over {-4 \rho  + 4 \rho^2 }} & {1\over {\rho  - \rho^2 }} & 
    -{1\over {4 \rho  - 4 \rho^2 }} \\
    {1\over {-4 \rho  + 4 \rho^2 }} & 
    {{1 + 2 \rho }\over {4 \rho  - 4 \rho^2 }} & 
    {{1 + 2 \rho }\over {4 \rho  - 4 \rho^2 }} & 
    {{1 + 2 \rho }\over {-4 \rho  + 4 \rho^2 }} & 
    {{1 + \rho }\over {2 \rho  - 2 \rho^2 }} & 
    -{1\over {4 \rho  - 4 \rho^2 }}  
  \end{array}
\right)
\end{equation}
\renewcommand{\arraystretch}{1}
with $\rho = \frac{k^2 q^2}{\kq^2}$. 

This choice has the advantage that the corresponding ``projectors''
$\{\hat{T}_{i, \mu \mus, \al \als}(q,k)\}$, which satisfy the
relations $\hat{T}_{i, \mu \mus, \al \als} \; T_j^{\mu \mus, \al \als}
= \delta_{ij}$, have straightforward forms: 
\begin{eqnarray}
\hat{T}_1^{\mu \mus, \al \als} &=& 
    \frac14
    \left( g^{\mu \mus} + 
         {{q^2 k^\mu k^\mus}\over 
           {\kq^2 - k^2 q^2}} \right)  
       \left( g^{\al\als} + 
         {{k^2 q^\al q^\als}\over 
           {\kq^2 - k^2 q^2}} \right) \,, \nonumber \\ 
\hat{T}_2^{\mu \mus, \al \als} &=& 
   \frac14
   \left( g^{\mu \mus} + 
         {{q^2 k^\mu k^\mus}\over 
           {\kq^2 - k^2 q^2}} \right)  
       \left( g^{\al \als} + 
         {{3 k^2 q^\al q^\als}\over 
           {\kq^2 - k^2 q^2}} \right) \,, \nonumber \\ 
\hat{T}_3^{\mu \mus, \al \als} &=& 
   \frac14
   \left( g^{\mu \mus} + 
         {{3 q^2 k^\mu k^\mus}\over 
           {\kq^2 - k^2 q^2}} \right)  
       \left( g^{\al \als} + 
         {{k^2 q^\al q^\als}\over 
           {\kq^2 - k^2 q^2}} \right) \,, \nonumber \\ 
\hat{T}_4^{\mu \mus, \al \als} &=& 
   \frac14
   \left( g^{\mu \mus} + 
         {{3 q^2 k^\mu k^\mus}\over 
           {\kq^2 - k^2 q^2}} \right)  
       \left( g^{\al \als} + 
         {{3 k^2 q^\al q^\als}\over 
           {\kq^2 - k^2 q^2}} \right) \,,\\ 
\hat{T}_5^{\mu \mus, \al \als} &=& 
   {k^2 q^2 \over 4 \kq [ \kq^2 - k^2 q^2 ] }
   \left( g^{\als \mus} k^\mu q^\al + 
          g^{\mu \als} k^\mus q^\al + 
          g^{\al \mus} k^\mu q^\als + 
          g^{\al \mu} k^\mus q^\als \right) \,, \nonumber \\
\hat{T}_6^{\mu \mus, \al \als} &=& 
   {k^2 q^2 \over 2 \kq^2} 
   \left( g^{\al \mus} g^{\mu \als} + 
          g^{\al \mu} g^{\als \mus} \right)\,. \nonumber 
\end{eqnarray}
We then write the decomposition 
\begin{equation}
  \label{eq:hdecomp}
   H^{\mu \mus, \al \als} = \sum_{i=1}^{6} H_i T_i^{\mu \mus, \al \als}\,,
\end{equation}
where the coefficients $H_i$ are
\begin{equation}
   H_i = \hat{T}_{i, \mu \mus, \al \als} \;  H^{\mu \mus, \al \als}.
\end{equation}
Since the $H_i$ depend only on $k$ and $q$, there exists a frame,
namely the photon-gluon centre of mass frame, in which the
quark-antiquark phase-space integration can be performed analytically
without difficulty. This would not have been the case if we had
already contracted with the photon and gluon projectors, as these
would have introduced a non-trivial dependence on an additional
external momentum (that of the proton). Our results for the
coefficients $H_i$ are:
\begin{eqnarray}
  H_{i} &=& \as e_Q^2 \left[ \Hbaro_{i} + 
     \Hbard_{i} {4 m^2 J^2 \over s k^2 q^2 + 4 m^2 J^2} + 
     \right. \nonumber \\ &+& \left.
     \Hbarl_{i} {J \over 4 \kq} \log {\kq + J \sqrt{1-{4 m^2\over s}}
       \over \kq - J \sqrt{1-{4 m^2\over s}}} \right] 
\end{eqnarray}
where $J = \sqrt{\kq^2-k^2 q^2}$ 
and
\begin{eqnarray}
\Hbaro_1 &=& {\sqrt{1-{4 m^2\over s}}\over J^2}  \left( -2 J^2 - k^2 s
  - q^2 s -  
         {{3 k^2 q^2 s^2}\over {4 J^2}} \right) \,, \nonumber\\ 
\Hbaro_2 &=& {\sqrt{1-{4 m^2\over s}}\over J^2}  \left( -2 J^2 - 2 k^2 s - 
         2 q^2 s - {{9 k^2 q^2 s^2}\over {4 J^2}}
          \right) \,, \nonumber\\ 
\Hbaro_3 &=& 
   {\sqrt{1-{4 m^2\over s}}\over J^2}  \left( -2 J^2 - 2 k^2 s - 2 q^2 s - 
         {{9 k^2 q^2 s^2}\over {4 J^2}} \right) \,, \nonumber\\ 
\Hbaro_4 &=& {\sqrt{1-{4 m^2\over s}}\over J^2}  \left( -2 J^2 - 3 k^2 s - 
         3 q^2 s - {{27 k^2 q^2 s^2}\over 
           {4 J^2}} \right)\,, \\ 
\Hbaro_5 &=& - {4 \sqrt{1-{4 m^2\over s}} k^2 q^2 s\over J^2 \kq} \,,
\nonumber\\  
\Hbaro_6 &=& - {8 \sqrt{1-{4 m^2\over s}} k^2 q^2\over \kq^2} \,,
\nonumber 
\end{eqnarray}
and 
\begin{eqnarray}
\Hbard_1 &=&  {\sqrt{1-{4 m^2\over s}}\over J^2} \left( J^2 - m^2 s -
  {k^2 q^2 s^2 \over 4 J^2} \right) \,, \nonumber\\  
\Hbard_2 &=& {\sqrt{1-{4 m^2\over s}}\over J^2} \left( J^2 - m^2 s +
  q^2 s - {k^2 q^2 s^2\over 4 J^2} \right) \,, \nonumber\\  
\Hbard_3 &=& {\sqrt{1-{4 m^2\over s}}\over J^2} \left( J^2 + k^2 s -
  m^2 s - {k^2 q^2 s^2 \over 4 J^2} \right) \,, \\ 
\Hbard_4 &=& {\sqrt{1-{4 m^2\over s}}\over J^2} \left( J^2 + k^2 s -
  m^2 s + q^2 s + {3 k^2 q^2 s^2\over 4 J^2} \right) \,, \nonumber\\  
\Hbard_5 &=& {\sqrt{1-{4 m^2\over s}} k^2 q^2 s^2 \over  J^4} \,,
\nonumber\\  
\Hbard_6 &=& {\sqrt{1-{4 m^2\over s}} k^2 q^2 \left( 2 J^2 - 4 m^2 s +
    s^2 \right) \over J^2 \kq^2} \,, \nonumber 
\end{eqnarray}
and
\begin{eqnarray}
\Hbarl_1 &=& {1\over 2 J^6} \Big[
       8 J^4 \kq^2 - 8 J^4 \kq s - 
       8 J^2 \kq k^2 q^2 s + 4 J^4 s^2 + \nonumber \\ &+&
       6 J^2 k^2 q^2 s^2 + 
       3 k^4 q^4 s^2 + 
       8 J^2 m^2 \left( -2 J^2 m^2 + 
          \kq^2 s \right) \Big] \,, \nonumber \\ 
\Hbarl_2 &=& {1\over 2 J^6} \Big[
       8 J^4 \kq^2 - 8 J^4 \kq s - 
       16 J^2 \kq k^2 q^2 s + 4 J^4 s^2 + \nonumber \\ &+&
       14 J^2 k^2 q^2 s^2 + 
       9 k^4 q^4 s^2 + 
       8 J^2 m^2 \left( 2 J^2 k^2 - 
          2 J^2 m^2 - J^2 s + 2 \kq^2 s \right) \Big] \,, \nonumber \\
\Hbarl_3 &=& {1\over 2 J^6} \Big[
       8 J^4 \kq^2 - 8 J^4 \kq s - 
       16 J^2 \kq k^2 q^2 s + 4 J^4 s^2 + \\ &+&
       14 J^2 k^2 q^2 s^2 + 
       9 k^4 q^4 s^2 + 
       8 J^2 m^2 \left( -2 J^2 m^2 + 
          2 J^2 q^2 - J^2 s + 2 \kq^2 s \right) \Big] \,, \nonumber \\
\Hbarl_4 &=& {1\over 2 J^6} \Big[
       8 J^4 \kq^2 - 8 J^4 \kq s - 
       24 J^2 \kq k^2 q^2 s + 4 J^4 s^2 +  \nonumber \\ &+&
       30 J^2 k^2 q^2 s^2 + 
       27 k^4 q^4 s^2 + 
       8 J^2 m^2 \left( -4 J^2 \kq - 
          2 J^2 m^2 + 3 \kq^2 s \right) \Big] \,, \nonumber \\ 
\Hbarl_5 &=& {8 k^2 q^2 s \over J^4} (\kq - m^2) \,, \nonumber \\ 
\Hbarl_6 &=& {8 k^2 q^2 \over J^2 \kq^2} 
      [ J^2 + 2 k^2 q^2 + \kq s  + 
       m^2 \left(s -4 \kq - 4 m^2 \right) ] \,. \nonumber
\end{eqnarray}
Though individually, each of the $\bar H$ suffer from a singularity at 
$J=0$, the results for the reduced cross sections are finite at $J=0$, 
and are given by:
\begin{equation}
  {\hat\sigma}_{F_i} \left(\xbf,\kt^2,Q^2,m^2 \right) = 
    {\as e_Q^2 \over 8} \xfb \kt^2 \bar{\sigma}_i
    \left(Q^2(\xfb-1)-\kt^2,-\kt^2,-Q^2,m^2\right) \,, 
\end{equation}
with
\begin{eqnarray}
\bar{\sigma}_i (s,k^2,q^2,m^2) &=& \left({q^2 \over 2 J^2}\right)^2 
          \left[ \; \ho_i + 
                 \hd_i  {4 m^2 J^2 \over s k^2 q^2 + 4m^2 J^2} + 
          \right.       \nonumber\\ &+&
          \left. (\hl_i + \hm_i) {J \over 4 \kq} 
  \log {\kq + J \sqrt{1-{4 m^2\over s}} \over \kq - J
    \sqrt{1-{4m^2\over s}}}  
 \; \right]\,,
\end{eqnarray}
where
\begin{eqnarray} 
\ho_2 &=& {\sqrt{1-{4 m^2\over s}} \over 2 J^4} \Big[ 
       -32 J^6 - 72 J^4 k^2 q^2 + 
        96 J^4 \kq s + \nonumber \\ && +
        360 J^2 \kq k^2 q^2 s - 48 J^4 s^2 - 
        330 J^2 k^2 q^2 s^2 - 
        315 k^4 q^4 s^2 \Big]\,, \nonumber
\\
\hd_2 &=& {\sqrt{1-{4 m^2\over s}} \over 2 J^4} \Big[ 
        16 J^6 + 24 J^4 k^2 q^2 - 
        32 J^4 \kq s - 16 J^4 m^2 s - 
        48 J^2 \kq k^2 q^2 s + \nonumber \\ && 
      - 36 J^2 k^2 m^2 q^2 s + 16 J^4 s^2 + 
        36 J^2 k^2 q^2 s^2 + 
        15 k^4 q^4 s^2 \Big] \,,
 \\
\hl_2 &=& {1\over J^6} \Big[
      32 {J^8} + 80 J^6 k^2 q^2 + 
      72 J^4 k^4 q^4 - 32 J^6 \kq s - 
      336 J^4 \kq k^2 q^2 s + \nonumber \\ && 
    - 360 J^2 \kq k^4 q^4 s + 
      16 J^6 s^2 + 240 J^4 k^2 q^2 s^2 + 
      540 J^2 k^4 q^4 s^2 + 
      315 k^6 q^6 s^2 \Big] \,,\nonumber
 \\
\hm_2 &=& {8m^2 \over J^4} \Big[ 
       -16 J^4 \kq - 8 J^4 m^2 - 
        36 J^2 \kq k^2 q^2 - 
        18 J^2 k^2 m^2 q^2 + 12 J^4 s + \nonumber \\ && +
        60 J^2 k^2 q^2 s + 
        45 k^4 q^4 s \Big]\,, \nonumber 
\end{eqnarray}
and
\begin{eqnarray} 
\ho_L &=& {\sqrt{1-{4 m^2\over s}} \over J^4} \Big[ 
       -24 J^4 k^2 q^2 + 
        16 J^4 \kq s + \nonumber \\ && +
        120 J^2 \kq k^2 q^2 s - 8 J^4 s^2 - 
        110 J^2 k^2 q^2 s^2 - 
        105 k^4 q^4 s^2 \Big] \,,\nonumber
 \\
\hd_L &=& {\sqrt{1-{4 m^2\over s}} k^2\over J^4}
      \Big[ 
        8 J^4 q^2 + 8 J^4 s + 
        8 J^2 k^2 q^2 s - 
        12 J^2 m^2 q^2 s + 8 J^2 q^4 s + \nonumber \\ && +
        8 J^2 q^2 s^2 + 5 k^2 q^4 s^2 \Big] \,,
 \\
\hl_L &=& {2 k^2 q^2 \over J^6} \Big[ 
        16 J^6 + 24 J^4 k^2 q^2 - 
        96 J^4 \kq s - 
        120 J^2 \kq k^2 q^2 s + \nonumber \\ && +
        72 J^4 s^2 +  
        180 J^2 k^2 q^2 s^2 + 
        105 k^4 q^4 s^2 \Big]\,, \nonumber 
 \\
\hm_L &=& {16 m^2 q^2 \over J^4} \Big[ 
        4 J^4 - 12 J^2 \kq k^2 - 
        6 J^2 k^2 m^2 + \nonumber \\ && +
        18 J^2 k^2 s + 
        15 k^4 q^2 s \Big] \,.\nonumber
\end{eqnarray}



\begin{thebibliography}{99}
\bibitem{DGLAP} V.N. Gribov and L.N. Lipatov, \sjnp{15}{1972}{438};\\
  G. Altarelli and G. Parisi, \npb{126}{1977}{298}; \\ Yu.L. Dokshitzer,
  \jetp{46}{1977}{641}. 
\bibitem{BFKL}  L.N. Lipatov, Sov. J. Phys. 23 (1976) 338;\\
       E.A. Kuraev, L.N. Lipatov and V.S. Fadin, Sov. Phys. JETP 45
       (1977) 199; \\
       Ya. Balitskii and L.N. Lipatov, Sov. J. Nucl. Phys. 28 (1978)
       822.
\bibitem{F2H1} H1 Collaboration (S. Aid et al.), \npb{470}{1996}{3}
  [\hepex{9603004}]. 
\bibitem{F2ZS} ZEUS Collaboration (M. Derrick et al.),
  \zpc{72}{1996}{399} [\hepex{9607002}].
\bibitem{MRST} A.D. Martin, R.G Roberts, W.J. Stirling and
  R.S. Thorne, \hepph{9805205}. 
\bibitem{GRV} M. Gluck, E. Reya and A. Vogt, \epjc{5}{1998}{461}
  [\hepph{9806404}]. 
\bibitem{BF} S. Forte and R.D. Ball, proceedings of
  \textit{International Workshop on Deep Inelastic Scattering and
    Related Phenomena (DIS 96)}, Rome, 1996, p.~208
  [\hepph{9607289}]. 
\bibitem{NLLBFKL}
  L.N. Lipatov and V.S. Fadin, \sjnp{50}{1989}{712}; \\
  V.S. Fadin, R. Fiore and M.I. Kotsky, \plb{539}{1995}{181}; \\ 
  V.S. Fadin, R. Fiore and M.I. Kotsky, \plb{387}{1996}{593}
  [\hepph{9605357}]; \\ 
  V.S. Fadin, and L.N. Lipatov, \npb{406}{1993}{259}; \\ V.S. Fadin,
  R. Fiore and A. Quartarolo, \prd{50}{1994}{5893} [\hepth{9405127}];
  \\ V.S. Fadin, 
  R. Fiore, and M.I. Kotsky, \plb{389}{1996}{737} [\hepph{9608229}];\\
  V.S. Fadin and L.N. Lipatov, \npb{477}{1996}{767} [\hepph{9602287}]; \\
  V.S. Fadin, M. I. Kotsky and L.N. Lipatov, \plb{415}{1997}{97}; \\
  S. Catani, M. Ciafaloni and F.Hautman, \plb{242}{1990}{97}; \\
  S. Catani, M. Ciafaloni and F.Hautman, \npb{366}{1991}{135}; \\
  V.S. Fadin, R. Fiore, A. Flashi, and M.I. Kotsky,
  \plb{422}{1998}{287} [\hepph{9711427}];\\ 
  V. Del Duca, \prd{54}{1996}{989};\\
  V. Del Duca, \prd{54}{1996}{4474};\\
  V. Del Duca and C.R. Schmidt, \prd{57}{1998}{4069} [\hepph{9711309}];\\
  V. Del Duca and C.R. Schmidt, \hepph{9810215}.
\bibitem{CCFL} M. Ciafaloni and G. Camici, \plb{430}{1998}{127}
  [\hepph{9803389}];\\ 
  M. Ciafaloni, \plb{429}{1998}{349} [\hepph{9801322}];\\
  M. Ciafaloni and G. Camici, \plb{412}{1997}{396} [\hepph{9707390}];\\
  V.S. Fadin and L.N. Lipatov, \plb{429}{1998}{127} [\hepph{9802290}].
\bibitem{Ross} D.A. Ross, \plb{431}{1998}{161} [\hepph{9804332}];\\
 E. Levin, \hepph{9806228}.
\bibitem{Bart} J. Bartels, H. Lotter and M. Vogt \plb{373}{1996}{215}.
\bibitem{Muel} A.H. Mueller, \plb{396}{1997}{251} [\hepph{9612251}].
\bibitem{CC1} G. Camici and M. Ciafaloni, \plb{395}{1997}{118}
  [\hepph{9612235}].
\bibitem{salam} G.P. Salam, \jhep{07}{1998}{019} [\hepph{9806482}].
\bibitem{BFKLQ2} R.K. Ellis, Z. Kunszt and E.M. Levin,
  \npb{420}{1994}{517};\\ 
  R.K. Ellis, F. Hautmann and B.R. Webber, \plb{348}{1995}{582}
  [\hepph{9501307}];\\ 
  R.D. Ball and S. Forte, \plb{351}{1995}{313} [\hepph{9501231}].
\bibitem{MuNa} A.H. Mueller and H. Navelet, \npb{282}{1987}{727}.
\bibitem{IR}
       A.H. Mueller, \plb{104}{1981}{161};\\
       B.I. Ermolaev and V.S. Fadin, \jetpl{33}{1981}{285};\\
       Yu.L.\ Dokshitzer, V.S. Fadin and V.A. Khoze, \zpc{15}{1982}{325};\\
       A.\ Bassetto, M.\ Ciafaloni and G.\ Marchesini and
       A.H. Mueller, \npb{207}{1982}{189};\\
       A.\ Bassetto, M.\ Ciafaloni and G.\ Marchesini,
       \prep{100}{1983}{201};\\
       Yu.L.\ Dokshitzer, V.A.\ Khoze, S.I.\ Troyan and A.H.\ Mueller,
       {\it Basics of Perturbative QCD}, Editions Frontieres, Paris,
       France 1991.
\bibitem{DKT} Yu.L. Dokshitzer, L.V. Gribov, V.A. Khoze and
  S.I. Troyan, \plb{202}{1988}{276};\\ 
L.V. Gribov, Yu.L. Dokshitzer, V.A. Khoze and S.I. Troyan,
\jetp{67}{1988}{1303};\\ 
M. Ciafaloni, A. Bassetto, G. Marchesini, \prep{100}{1983}{201}.
\bibitem{CCFM}   M. Ciafaloni, \npb{296}{1987}{249};\\
  S. Catani, F. Fiorani and G. Marchesini, \plb{234}{1990}{339};\\
  S. Catani, F. Fiorani and G. Marchesini, \npb{336}{1990}{18}.
\bibitem{GM} G. Marchesini, \npb{445}{1995}{40} [\hepph{9412327}].
\bibitem{BMSS} G. Bottazzi, G. Marchesini, G.P. Salam and
  M. Scorletti, \npb{505}{1997}{366} [\hepph{9702418}].
\bibitem{RyskPriv} M. Ryskin, private communication.
\bibitem{OrrStir} L.H. Orr and W.J. Stirling, \prd{56}{1997}{5875}
  [\hepph{9804331}];\\ 
  L.H. Orr and W.J. Stirling, \plb{436}{1998}{372} [\hepph{9806371}].
\bibitem{FSv} J.R. Forshaw and A. Sabio Vera, \plb{440}{1998}{141}
  [\hepph{9806394}]. 
\bibitem{Webb} B.R. Webber, \plb{444}{1998}{81} [\hepph{9810286}].
\bibitem{GPSfuture} G.P. Salam, in preparation.
\bibitem{CMWfuture} S. Catani,G. Marchesini and B.R. Webber, in
  preparation. 
\bibitem{H1Et} H1 collaboration, contrib.\ paper pa02-073 to ICHEP'96,
  Warsaw 1996.
\bibitem{H1kts} H1 collaboration (C. Adloff et al.),
  \npb{485}{1997}{3} [\hepex{9610006}].
\bibitem{ZEUSfj} ZEUS Collaboration (J. Breitweg  et al.),
  \hepex{9805016}. 
\bibitem{H1fj96} H1 Collaboration (C. Adloff et al.),
  \hepex{9809028}. 
\bibitem{KMSccfm} J. Kwiecinski, A.D. Martin and P.J. Sutton,
  \prd{53}{1996}{6094} [\hepph{9511263}].
\bibitem{SMALLX} G. Marchesini and B.R. Webber, \npb{386}{1992}{215}.
\bibitem{Biernat} K. Golec-Biernat, L. Goerlich and J. Turnau,
  \npb{527}{1998}{289}  [\hepph{9712345}].
\bibitem{LDCorig} B. Andersson, G. Gustafson, J. Samuelsson,
  \npb{467}{1996}{443}. 
\bibitem{LDCmc}  H. Kharraziha and L. L\"onnblad, \jhep{03}{1998}{006}
  [\hepph{9709424}]. 
  \prd{53}{1996}{6094} [\hepph{9511263}].
\bibitem{CCH} S. Catani, M. Ciafaloni and F. Hautmann, \npb{366}{1991}{135}.
\bibitem{CollinsEllis} J.C. Collins and R.K. Ellis, \npb{360}{1991}{3}.
\bibitem{LevinEtAl} E.M.Levin, M.G. Ryskin, Yu.M. Shabelskii and
  A.G. Shuvaev, \sjnp{53}{1991}{657}.
\bibitem{BW} J. Bartels, V. Del Duca, A. De Roeck, D. Graudenz,
  and M. W\"usthoff,  \plb{384}{1996}{300};\\
  J. Bartels, V. Del Duca and M. W\"usthoff, \zpc{76}{1997}{75}.
\bibitem{KLM} J. Kwiecinski, S.C. Lang and A.D. Martin,
  \epjc{6}{1999}{671} [\hepph{9707240}]. 

\bibitem{F2cH194} H1 Collaboration (Adloff et al.),
  \zpc{72}{1996}{593} [\hepex{9607012}].
\bibitem{F2cZS94} ZEUS Collaboration (Breitweg et al.)
  \plb{407}{1997}{402} [\hepex{9706009}]. 
\bibitem{NPRW} H. Navelet, R. Peschanski, C. Royon and S. Wallon,
  \plb{385}{1996}{357}, [\hepph{9605389}].
\bibitem{DW}
 Yu.L.\ Dokshitzer and  B.R.\ Webber,  \plb{404}{1997}{321}
 [\hepph{9704298}]. 
\bibitem{frag} J. Binnewies, B.A. Kniehl and G. Kramer,
  \prd{52}{1995}{4947} [\hepph{9503464}];\\
  J. Binnewies, B.A. Kniehl and G. Kramer, \prd{53}{1996}{3573}
  [\hepph{9506437}]. 
\bibitem{LEPTO} G. Ingelman, A. Edin and J. Rathsman,
  \cpc{101}{1997}{108} [\hepph{9605286}].
\bibitem{MuFJ} A.H. Mueller,  \npps{18 C}{1990}{125};\\
  A.H. Mueller,  \jpg{17}{1991}{1443}.
\bibitem{DISFJ} J. Bartels, A. De Roeck and M. Loewe,
  \zpc{54}{1992}{635};\\
  J. Kwiecinski, A.D. Martin and P.J. Sutton, \plb{287}{1992}{254};
  \prd{46}{1992}{921};\\
  W.K. Tang, \plb{278}{1992}{363};\\
  J. Bartels, M. Besancon, A. De Roeck and J. Kurzhoefer, \textit
    {proceedings of the HERA Workshop 1992} (eds.\ W. Buchm\"uller and
    G. Ingelman), p.~203.
\bibitem{MEPJET} E. Mirkes and D. Zeppenfeld, \plb{380}{1996}{205}
  [\hepph{9511448}];\\ 
  T. Brodkorb and E. Mirkes, \zpc{66}{1995}{141} [\hepph{9402362}].
\bibitem{MathHIP} A. Hsieh and E. Yehudai, \cphys{6}{1992}{253}.


\end{thebibliography}
\end{document}